\documentclass[12pt]{article}

\usepackage{array,epsfig}
\usepackage{subcaption}
\usepackage{amsmath}
\usepackage{amsfonts}
\usepackage{amssymb}
\usepackage{amsxtra}
\usepackage{amsthm}
\usepackage{mathrsfs}
\usepackage{color}
\usepackage{fancyhdr}
\usepackage{setspace}
\usepackage{xcolor}
\usepackage[authoryear]{natbib}
\RequirePackage[colorlinks,citecolor=blue,urlcolor=blue]{hyperref}
\RequirePackage{graphicx}

\theoremstyle{definition}

\newcommand{\bb}[1]{\mathbb{#1}}
\newcommand{\mcl}[1]{\mathcal{#1}}
\newcommand{\mbf}[1]{\mathbf{#1}}

\newcommand{\Y}{\mbf{Y}}

\newcommand{\T}{\mbf{T}}
\newcommand{\cov}{\text{Cov}}

\newcommand{\one}{\mathbf{1}}

\newcommand{\s}{\mathbf{s}}
\newcommand{\vect}[1]{\text{vec}(#1)}

\newcommand{\IG}{\text{Inverse-Gamma}}
\def\iidsim{~{\buildrel \text{iid}\over \sim}~}

\newcommand{\pink}[1]{\textcolor{black}{#1}}
\newcommand{\green}[1]{\textcolor{black}{#1}}
\newcommand{\red}[1]{\textcolor{black}{#1}}

\providecommand{\keywords}[1]
{
  \textbf{\textit{Keywords---}} #1
}
\title{
Improved fMRI-based Pain Prediction using Bayesian Group-wise Functional Registration
}
\author{
 {Guoqing Wang, Abhirup Datta, Martin A. Lindquist}\\
 {Department of Biostatistics,
 Johns Hopkins University,
 Baltimore, MD}
 }

\begin{document}
\doublespace
\maketitle
\begin{abstract}
In recent years, the field of neuroimaging has undergone a paradigm shift, moving away from the traditional brain mapping approach towards the development of integrated, multivariate brain models that can predict categories of mental events. However, large inter-individual differences in both brain anatomy and functional localization \emph{after} standard anatomical alignment remain a major limitation in performing this type of analysis, as it leads to feature misalignment across subjects in subsequent predictive models. This paper addresses this problem by developing and validating a new computational technique for reducing misalignment across individuals in functional brain systems by spatially transforming each subject’s functional data to a common latent template map. Our proposed Bayesian functional group-wise registration approach allows us to assess differences in brain function across subjects and individual differences in activation topology. We achieve the probabilistic registration with inverse-consistency by utilizing the generalized Bayes framework with a loss function for the symmetric group-wise registration. It models the latent template with a Gaussian process, which helps capture spatial features in the template, producing a more precise estimation. We evaluate the method in simulation studies and apply it to data from an fMRI study of thermal pain, with the goal of using functional brain activity to predict physical pain. We find that the proposed approach allows for improved prediction of reported pain scores over conventional approaches.
\end{abstract}


\newpage

\section{Introduction}


Functional magnetic resonance imaging (fMRI) is one of the most widely used methods for investigating human brain function and has provided critical insight into our understanding of human behavior. Traditional brain mapping approaches treated the brain as the outcome and task, or behavioral variables, as predictors. Here the goal was to assess, at each spatial location, whether or not there is a non-zero effect related to the task. Recently, analysis has shifted towards a predictive modeling approach that uses statistical learning to develop integrated models of activity across multiple brain regions (i.e., brain signatures) to predict clinical outcomes \citep{orru2012using, haynes2015primer}. However, a major limitation holding back these types of models is large inter-individual differences in functional brain topology, as this can lead to a situation where spatial locations are misaligned across subjects, thereby reducing the ability to accurately predict outcomes. We elaborate on this problem and our proposed solution below.


During a task-based fMRI study, subjects are administered one or more stimuli while their brain is scanned at hundreds of time points. At each time point, the subject’s blood oxygenation level dependent (BOLD) response is measured at a large number of spatial locations (referred to as voxels), giving rise to multivariate time series data. 
To perform population-level statistical analysis, voxels need to consistently lie within the same brain structure across subjects. Current standard practice towards preprocessing fMRI data includes using nonlinear transformations to transform individual subjects’ anatomical data to an average, anatomically based 3D reference space (for example, ``Montreal Neurologic Institute'' (MNI) space). Those transformations are then applied to functional data collected during the same scanning session, allowing the resulting data to be compared across subjects \citep{lindquist2008statistical, ombao2016handbook}. 

While this approach ensures the brains of all subjects overlap with regards to gross anatomical landmarks, it does not consider residual differences in brain anatomy, or the location and distribution of functional regions relative to these landmarks.  This is problematic as a large number of statistical analyses are performed voxel-wise (e.g., subjects are compared using a separate model at each voxel). Inter-subject variability in functional topology can lead to a situation where voxels from different functional brain regions are compared to one another in a model, significantly complicating interpretation. In addition, voxels are often used as features in predictive models, and any variation in functional location across subjects will lead to feature misalignment when training models. This points to the need to map individual brains into a functional population-level template space after the initial alignment to an anatomically based brain space and prior to subsequent statistical modeling.


One proposed solution to this problem is the so-called ‘hyperalignment’ procedure \citep{Haxby2011}, where brain activity patterns corresponding to cognitive events are represented as vectors in a neural representational space spanned by the voxels in a local neighborhood. Each participant’s local voxel-wise activity pattern is rotated in multivariate voxel space using a Procrustes transformation to align the representational geometry across subjects. This body of research includes a number of extensions, including: kernel hyperalignment \citep{Lorbert2012}, regularized hyperalignment \citep{Xu2012}, the two-phase joint SVD-Hyperalignment algorithm \citep{Chen2014}, the shared response model \citep{Chen2014}, and searchlight hyperalignment \citep{Guntupalli2016}. In addition, other methods such as functional time series alignment \citep{Sabuncu2010} and functional connectivity alignment \citep{Conroy2013} have been developed that align subjects based on inter-subject correlations in time-series data during movie viewing. 

Hyperalignment procedures show great promise for allowing participants to vary in their functional activation patterns \citep{Haxby2011}, but suffer from several shortcomings. First, hyperalignment requires subjects to watch a long film (up to 2 hours) to align subjects into a common representational space, and functional connectivity-based methods require substantial resting state fMRI data. Second, the choice of template data influences which types of functional patterns can be appropriately aligned. Third, these methods do not include an explicit spatial model able to make inferences on the location or extent of activation. 

\citet{wang2021bayesian} proposed to address these concerns by introducing a Bayesian model for registering local information contained in an fMRI activation map to a pre-specified template map in a pair-wise manner.  
The method quantified the uncertainty of the transformation by estimating the posterior distribution of transformation functions. The template map was selected {\it a priori} based on scientific knowledge. This approach has computational advantages, as it is a variation of pair-wise registration that can easily be performed in parallel across subjects. However, a major shortcoming of the approach was the need to pre-specify the template map, which could potentially bias estimation.

A group-wise extension of this registration approach would explicitly address this selection bias by allowing the template map to be a latent variable and estimating it in a joint analysis for simultaneously registering all the images. \citet{joshi2004unbiased} previously proposed such an approach. Their approach alternated between the following two steps until convergence: 1) the set of images are all transformed to a common template; and 2) the template is updated by averaging the transformed images. This approach has a drawback in that it may lead to a fuzzy template in the case where the initial input images are noisy. In addition, \citet{allassonniere2007towards} criticized this approach due to its lack of probabilistic interpretation, and proposed a generative model for the template. Here the template was parameterized as the linear combination of penalized radial basis functions. 
However, as shown in our simulation study, this approach underestimates the variability at locations with less supportive information. 
\citet{zhang2013bayesian} proposed an extension to the registration problem using large deformation diffeomorphic metric mappings (LDDMM) to estimate the deformation fields and the latent template. Extensive priors were assigned on the deformations, and a noninformative prior was assigned to the template. In their work, the template was updated by averaging across the transformed images, without considering spatial variation.  In this work we consider the work by \citeauthor{allassonniere2007towards} as the conventional approach to which we compare our proposed algorithm, as both approaches explicitly model the spatial features of the latent template. 

A desirable property when performing image registration is inverse-consistency, which describes the ability of a model to simultaneously register subject-specific floating maps to the common template map and the template to the floating maps \citep{Inverse-consistency1, Inverse-consistency2}. The generative registration models considered in \citeauthor{allassonniere2007towards} does not incorporate inverse consistency, as its likelihood function involves the registration from the template map to the floating maps, but not the opposite registration direction. This is a compromise made to obtain probabilistic interpretation.
However, \cite{Inverse-consistency1} showed that estimating the forward and backward transformations independently might lead to an inconsistent correspondence between maps.  They further tackled  inverse-consistency through optimization by introducing a loss function for symmetric registration that simultaneously performs registration in both directions. To the best of our knowledge, probabilistic modeling in the area of symmetric group-wise registration has not previously been explored. 


We propose to address inverse consistency by proposing a symmetric group-wise registration within a probabilistic framework. This is achieved by using a {\em generalized Bayes framework}, which is a flexible loss-function based pseudo-likelihood approach to valid Bayesian inference that is gaining increased traction due to its robustness to model misspecification and other desirable properties \citep{chernozhukov2003mcmc,bissiri2016general}. The pseudo-likelihood is a representation of the inverse-consistency loss that quantifies the registrations between the floating maps and the latent template map. In our application, we choose bi-directional sum-of-squared difference (SSD) as our loss function, though it can be replaced by other similarity measures depending on the context. 

Our approach models the latent template with a Gaussian process, which helps to capture spatial features and produce less noisy template estimation. The fully Bayesian probabilistic model enables inference on the latent map via posterior samples. One of the major limitations of the Gaussian process model is the computational difficulty, especially when dealing with a large number of spatial locations. We therefore employ the nearest-neighbor Gaussian process (NNGP) to enhance computational efficiency \citep{datta2016hierarchical}. 
The NNGP model deploys nearest neighbor graphical models to capture spatial correlation, and provides highly scalable and fully model-based inference for large spatial datasets. However, standard applications of NNGP necessitates a fixed design for the spatial locations. In our setting, the locations (transformed voxel locations) change with each update of the transformation parameters in the MCMC. 
We develop a variant of NNGP for the group-wise registration problem, that can handle such dynamic designs while remaining computationally scalable. 

Our methodological contributions in this work can be summarized as follows: (1) providing a probabilistic model to the symmetric group-wise registration approach conforming to inverse-consistency using the generalized Bayes framework; (2) offering the ability to make inferences on the template map through the the posterior distribution; and (3) incorporating Gaussian processes for spatially explicit group-wise registration and providing an computationally efficient implementation scheme. 

This work is motivated by an fMRI study of thermal pain. Physical pain is associated with enormous social and economic costs \citep{simon2012relieving}. However, pain can be difficult to determine as it is primarily assessed by means of self-report. The ability to accurately self-report pain is limited in many vulnerable populations, and self-reports provide a limited basis for understanding the neurophysiological processes underlying pain. Recent work has attempted to use predictive modeling approaches to derive patterns of activity across brain regions that can be used to provide direct measures of pain intensity \citep{wager2013fmri, lindquist2017group}. To be clinically useful these models need to exhibit a high degree of sensitivity and specificity to pain outcomes. Here inter-individual differences in functional brain topology are critically holding back the ability for these models to achieve their full potential.

In the study, participants experienced multiple levels of painful heat while undergoing fMRI scanning. The thermal stimuli were administered at six different temperature levels, and the participants reported the corresponding pain intensity on a 100-point scale.  The within-subject functional activation maps associated with each stimuli are used as features in a predictive model designed to predict reported pain. These maps are the output of a series of voxel-wise models that quantify the activation in the BOLD signal that arises due to the stimulus of interest. They correspond to spatial maps of the regression coefficient related to the predictor modeling the stimulus. These models are generally fit after an initial anatomical registration has already been performed. Thus, all subject-specific activation maps should be in a common anatomical space, and all voxel-specific values directly comparable across participants. However, as discussed above, there remain substantial individual variation in functional brain anatomy even after anatomical registration that causes this assumption to be invalid. To circumvent this issue, we seek to perform a secondary functional registration of the data into a common functional space via our groupwise registration prior to performing subsequent statistical analysis. Our hypothesis is that functional registration will help improve the sensitivity and specificity of pain prediction over equivalent approaches that do not perform functional alignment.

This paper is organized as follows. In Section \ref{Section: Pain data description} we provide details about the data from an fMRI study of thermal pain. Section \ref{Section: Review} reviews the class of conventional probabilistic models for group-wise registration. We introduce our probabilistic symmetric group-wise registration using generalized Bayes in Section \ref{section:full model}. In Section \ref{Section: Simulation} we present simulation studies for both one-dimensional curve registration and two-dimensional image registration. In Section \ref{Section: Application to the pain data} we apply the method to the pain data and demonstrate how images transformed by our approach yield increased capability to predict pain, than the standard images. The paper concludes with a Discussion. 

\section{Thermal Pain Study}\label{Section: Pain data description}
The data were acquired from an fMRI study of thermal pain; see \cite{woo2015distinct} for an in-depth discussion of the data set. In brief, $33$ healthy, right-handed subjects completed the study (age $27.9 \pm 9.0$ years, $22$ females). All subjects gave informed consent, and the Columbia University Institutional Review Board approved the study. 
\green{For each subject, between $58 - 75$ stimuli were administered for $12.5 \, s$ at six different temperatures, ranging from $44.3^{\circ}$ to $49.3^{\circ}$ C. Each stimulus was followed by a $4.5 - 8.5$ second rest period, after which subjects rated their intensity of pain on a scale between $0$ and $100$. Each trial ended with a $5 - 9$ second rest period, followed by a new stimulus.}

\green{While the study was being performed, whole-brain fMRI data were acquired on all participants using a 3T Philips Achieva TX scanner. Structural images were acquired using high-resolution T1 spoiled gradient recall images (SPGR). For each subject, 1845 functional EPI images were acquired with TR$=2000 \, ms$, TE$=20 \, ms$, field of view$=224\, mm$, $64 \times 64$ matrix, $3 \times 3 \times 3 \, mm^3$ voxels, 42 interleaved slices, parallel imaging, and SENSE factor 1.5.  The structural images were coregistered to the mean functional image using the iterative mutual information-based algorithm in SPM8\footnote{Statistical Parametric Mapping, version 8; http://www.fil.ion.ucl.ac.uk/spm/}. The images were then normalized to Montreal Neurological Institute (MNI) space using SPM8’s generative segment-and-normalize algorithm.}

For each subject, a voxel-wise general linear model (GLM) analysis was performed using SPM8. 
\green{Here the fMRI time series at the voxel of interest was the outcome in the model. To create regressors for each temperature level, boxcar functions corresponding to the the $12.5 \, s$ thermal stimulation were convolved with a canonical hemodynamic response function \citep{lindquist2009modeling}. A number of additional  regressors of non-interest (i.e., nuisance variables) were added, including: a run specific intercept; linear drift across time within each run; the six estimated head movement parameters, their mean-centered squares, their derivatives, and squared derivative for each run; indicator vectors for outlier time points identified based on their multivariate distance from the other images in the sample; indicator vectors for the first two images in each run; and the average signal from white matter and the ventricles. }
For each temperature the estimated regression coefficients corresponding to that temperature specific regressor at each voxel were combined into a single activation map, resulting in six different maps depicting the functional response across the brain in reaction to each temperature. Further details of the prepossessing and analysis steps can be found in \cite{wang2021bayesian}. 

\begin{figure}
    \centering
    \includegraphics[width=\textwidth]{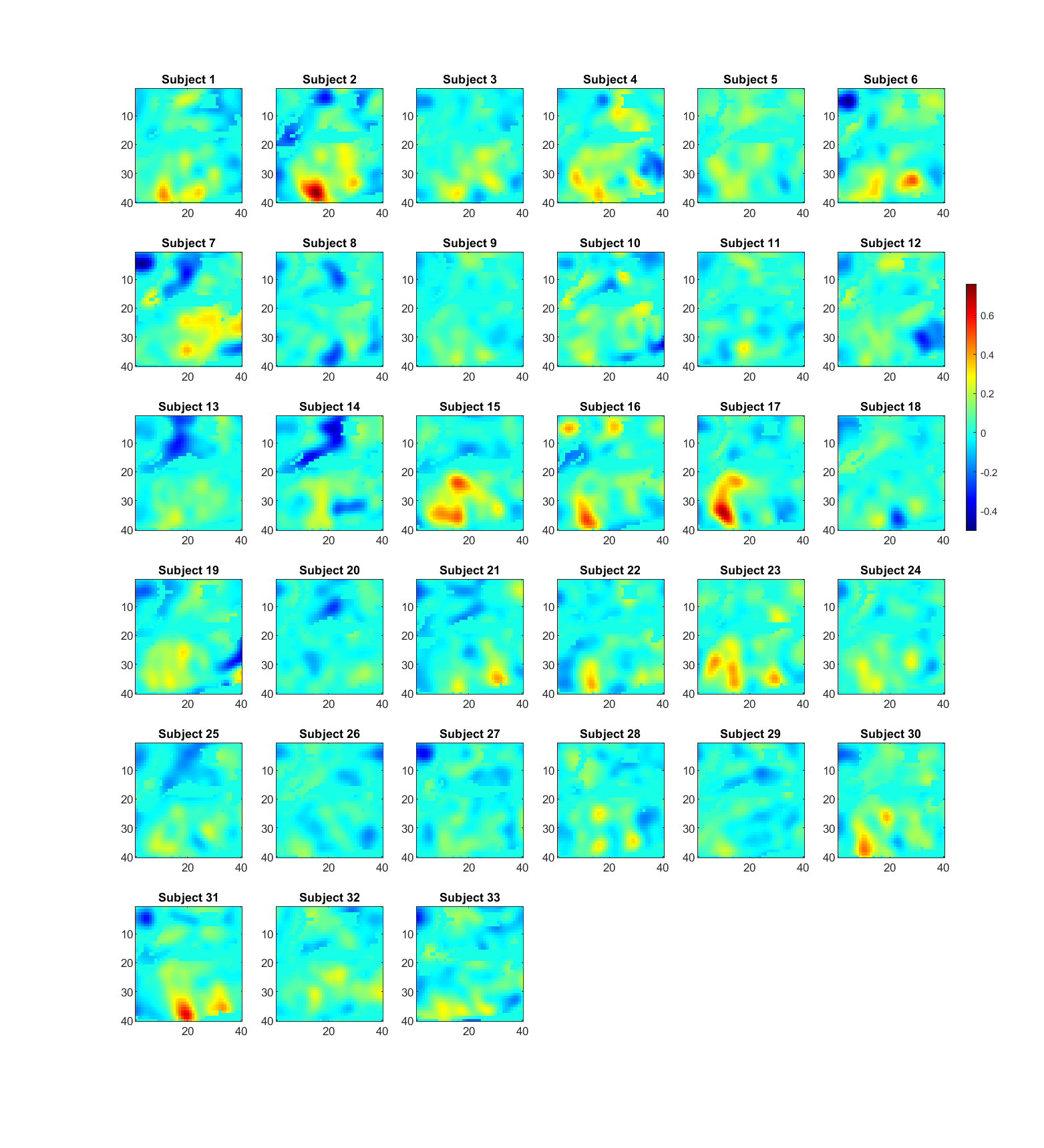}
    \caption{Subject-specific mean functional activation maps for 33 subjects averaged across six different thermal stimulus levels.}
    \label{fig: real Raw}
\end{figure}

\green{Figure \ref{fig: real Raw} shows the subject-specific mean functional activation maps across the six different temperatures for all 33 subjects within a pre-specified circular region of the brain with radius 15 voxels. In this region one can identify several local peaks within individual participants that conform to known somatosensory areas, which are located in the bottom part of each map. However, differences in spatial location of the peaks still persist between subjects. For example, the most active voxels for subject 1 is located towards the left corner of the activation area, while it is roughly in the middle bottom region for subject 2. 
We focus our analysis on a somatosensory region of the brain, as it is known to be related to pain \citep{bushnell1999pain, gustin2012pain, vierck2013role}. It is reliably observed in pain studies as evidenced by meta-analyses of neuroimging pain studies; see, for example, \cite{yarkoni2011large}. }

\section{Probabilistic models for the registration problem}\label{Section: Review}
We begin by briefly summarizing the probabilistic model first introduced by \cite{allassonniere2007towards}. Let $Y_i(s_v)$ be the intensity value in the spatial map of subject $i$ at location $s_v$ corresponding to voxel $v$, for $i\in\{1,\cdots, N\}$ and $v\in \{1, \cdots V\}$. Let $\Y = \{Y_1, Y_2, \ldots, Y_N\}$ be the collection of the subject-specific brain activation maps $Y_i = \{Y_i(s_v): v=1,\ldots,V\}$. The latent activation map to which we seek to register the maps is denoted $X = \{X(s_v): v\in \{1, \cdots V\} \}$.
Further, they assume an unobserved subject-wise deformation field $z_i:\mathbb{R}^2\rightarrow\mathbb{R}^2$ for $i$ such that 
\begin{equation}\label{eq:gen}
    Y_i(s_v) = X(T_i(s_v))+\sigma_i\epsilon,
\end{equation} where $T_i(s_v) = s_v-z_i(s_v)$ and $\epsilon\iidsim{N(0,1)}$. Here the parameter $\sigma_i$ characterizes the subject-specific noise. 

The latent activation map $X$ is assumed to belong to a reproducing kernel Hilbert space $\mathbb{V}$ which is determined by the kernel $\mathbf{K}$. The model is defined on a fixed finite dimensional subspace of $\mathbb{V}$ with a set of landmark points $\tilde s_p$ for $1\leq p \leq P$. The choice of landmarks is a trade-off between the desired accuracy of estimation and computational complexity. $X$ is parameterized by the linear combination of the kernels centered at the landmarks with weights $w\in\mathbb{R}^P$, i.e.,
\begin{align}
    X(s)=\sum_{p=1}^P K(s, \tilde s_p) w(p), \;\;\text{for}\; s \in \{s_1, \ldots, s_V\}.
\end{align}
Priors on the weights, $w\sim N(0, \mbf{K})$ with $\mbf{K}_{i,j} = K(\tilde s_i, \tilde s_j)$ for $1\leq i, j\leq P$, 
are equivalent to the priors on the latent template, which can be seen a penalization on the norm of $\bb{V}$. The choice of the prior distribution on the deformation field (i.e. transformation function) $T$ depends on its format, denoted $p(T)$. For instance, \citeauthor{allassonniere2007towards} focused on non-parametric deformation fields, and parameterized the transformations as a linear function of kernels. Similar to $X$, the prior distribution of $T$ is placed on the weights. One can also consider an affine transformation, which we will describe in detail in the following section. The inverse-gamma distribution is assigned as the prior distribution of $\sigma_i^2$, i.e., $\sigma_i^2\sim \IG(a_0, b_0)$. 
A common choice for the kernel function is a radial Gaussian kernel, $K(s, \tilde s_p) = \exp{(-\|s-\tilde s_p\|^2/\tau^2)}$ for locations $s$ and $\tilde s_p$.

Though the original work proposed a framework to find the maximum a posteriori probability (MAP) estimate using the Expectation-Maximization (EM) algorithm, one can also sample through Markov Chain Monte Carlo (MCMC) to approximate the posterior distribution of the latent map and the transformation functions.

\section{Symmetric Group-wise Registration with Gaussian Processes}\label{section:full model}
The model described in the previous section provides a way to transform the latent map onto each subject's activation map. However, this generative model is one-directional and it is infeasible to obtain legitimate updates on the inverse transformation from the activation maps to the latent template. As a consequence, it is less beneficial in the context of fMRI data analysis, as we are typically interested in conducting group-level inference on the transformed activation maps. To accommodate this, we employ a symmetric group-wise registration model, which considers registration in both directions simultaneously. In this section, we provide details for performing symmetric group-wise registration of the functional activation maps using a Gaussian process model.

The registration model described in Section \ref{Section: Review},  inherently uses the sum of squared differences (SSD), $\|Y_i-\beta_i X(T_i)\|^2_{L^2}$, as the loss function coming from the negative log-likelihood of the generative model (\ref{eq:gen}). Here $\beta_i\in \bb{R}^+$ is a scaling-correction parameter for each subject, so that the subject-specific images are on the same intensity-level of the scaling-corrected transformed template, $\beta_i X(T_i)$. We introduce the $\beta_i$'s as an alternative to apriori standardizing the images onto the same intensity level. The collection of scaling-correction parameters is denoted by $\mbf{\beta} = \{\beta_1, \beta_2, \ldots, \beta_N\}$.

We extend this squared error loss function to a symmetric loss that accounts for registration $T_i$ both from the latent to individual subject activation map, and a registration $T_i^{r}$ in the reverse direction, i.e., from the each individual activation map to the latent template. Let $\mbf{T} = \{T_1, T_2, \ldots, T_N\}$ and $\mbf{T}^r = \{T_1^r, T_2^r, \ldots, T_N^r\}$ be the collection of transformations. The symmetric loss function can be expressed using a bi-directional sum-of-squares-differences as follows:
\begin{equation}   \label{eqn: symmetric loss function}
\begin{aligned}
   \ell_{sym}(X,\mbf{T},\mbf{T}^r, \mbf{b}, \mbf{\sigma})=\sum_{i=1}^N \bigg\{&\frac{1}{\sigma_i^2}\|Y_i(T_i^{r})- \beta_i X\|^2_{L^2}+\frac{1}{\sigma_i^2}\|Y_i-\beta_iX(T_i)\|^2_{L^2}\\
   &+ \lambda^{r}\|T_i(T_i^{r}(\cdot)) -Id(\cdot)\|_F+ \lambda^{r}\|T_i^{r}(T_i(\cdot)) -Id(\cdot)\|_F\bigg\},
\end{aligned}
\end{equation}
where $\|\cdot\|_F$ is the Frobenius norm, $X(T_i)$ is the latent map transformed by the forward transformation $T_i$.  $Y_i(T_i^{r})$ is the subject-specific map transformed by the backward transformation $T_i^{r}$. \pink{Given $T_i^r$, we employ the cubic interpolation method for $Y_i(T_i^{r})$}.  The latter two regularization terms penalize the extreme case that the forward and backward transformations are far away from being the inverses of each other. Two penalization terms are needed owing to the noncommutativity of general affine transformations, i.e., $T_i(T_i^{r}(\cdot)) \neq T_i^{r}(T_i(\cdot))$. The hyperparameter $\lambda^r$ balances the forward and backward transformations, and $Id(\cdot)$ is the identity transformation. We utilize the Watanabe-Akaike information (WAIC; \cite{Watanabe2010}) to choose $\lambda^r$.

Given a prior distribution on the unknown quantities $(X,\T,\T^r,\mbf{\beta},\mbf{\sigma})$, 
the {\em generalized posterior} or {\em Gibbs posterior} distribution given the symmetric loss function above will be given by 
\begin{align}\label{eqn: symmetric posterior}
    p(X,\T,\T^r, \mbf{\beta}, \mbf{\sigma}|\Y)\propto\exp{\big\{-\ell_{sym}(X,\T,\T^r,\mbf{\beta},\mbf{\sigma})\big\}} p(X,\T,\T^r,\mbf{\beta},\mbf{\sigma}), 
\end{align} assuming some normalizing constants. Unlike the conventional model in Section \ref{Section: Review}, the term $\exp{\{-\ell_{sym}(X,\T,\T^r,\mbf{\beta},\mbf{\sigma})\}}$ in (\ref{eqn: symmetric posterior}) does not correspond to any generative model for the data as the loss function (\ref{eqn: symmetric loss function}) is not derived as a negative log-likelihood. Hence the Gibbs posterior is not a true posterior in the traditional Bayesian sense. However, \cite{chernozhukov2003mcmc} has established extensive desirable large sample properties of such generalized Bayes updates.
Indeed, \citet{bissiri2016general} argued that updates of the form (\ref{eqn: symmetric posterior}) are the only coherent update of a prior belief $p(\cdot)$ given the data and the choice of the loss function $\ell$. Thus, the use of Gibbs posteriors are becoming an increasingly common form of Bayesian inference, termed as generalized Bayes \citep{grunwald2020fast,fiksel2021generalized,rigon2020generalized}. In our case, adopting a generalized Bayes framework with the symmetric loss allows probabilistic inference for group-wise registration under inverse consistency. The general framework outlined here is agnostic to the choice of loss function, and other choices such as mutual information loss \citep{Viola1997, Thevenaz1998} and correlation-based losses \citep{Gruen1987}, can replace the squared error losses depending on the application. 

\subsection{Prior Distributions}\label{section: prior distributions}
We assume a Gaussian process (GP) prior for the latent template $X$ to capture its spatial features. On the set of voxels $\{1,\ldots,V\}$, this implies that $X$ is a multivariate normal distribution with mean $0$ and covariance matrix $\Sigma \in \mathbb{R}^{V\times V}$. Here we use the popular exponential covariance kernel  $\Sigma_{i,j} = C(s_i, s_j) = \alpha \exp{(-\rho\|s_i - s_j\|)}$, where $\alpha$ denotes the spatial variance and $\rho$ is the spatial decay parameter. With the transformation $T_i$ for subject $i$, the conditional distribution of the transformed latent map $X(T_i)$ given the latent map $X$ can be derived through Kriging \citep{banerjee2014hierarchical, cressie2015statistics}, which provides the conditional distribution at the spatially transformed locations $T_i(\{1,\ldots,V\})$. We summarize the priors as follows, 
\begin{equation}\label{eq: priors of X and X(T_i)}  
\begin{aligned}
    &X(T_i)|X, T_i\sim N(P_i X,\; S_i), \\
    &X \sim N(0, \Sigma),
\end{aligned}
\end{equation}
where $P_i = C_{i}^T\Sigma^{-1}$ and $S_i = \Sigma_{T_i} - P_i \Sigma P_i^T$. Here $\Sigma_{T_i}$ is the marginal GP covariance matrix of $X(T_i)$ with the $(j,k)^{th}$ entry defined as $\alpha\exp{(-\rho\|T_i(s_j)-T_i(s_k)\|)}$. 
$C_{i} = \cov(X(T_i), X) = \alpha \exp(-\rho\|T_i(s_j) - s_k\|)$ defines the covariance between $X(T_i)$ and $X$. 
\red{Note that there is no conditional distribution for $Y_i(T_i^r)$, as we are considering it as data given $T_i^r$, which can be gotten by interpolation.}
The prior distributions of the transformations $T_i$ and $T_i^r$ act as a regularization term in the registration problem which is inherently ill-posed \citep{Fischer2003}. We here adopt an idea introduced by \citeauthor{wang2021bayesian} to initiate the prior distributions. Assume an affine transformation, 
\begin{equation}
T(s) = As+b, 
\end{equation}
where $s = (s_x, s_y)^T$, $A=\begin{pmatrix}
A_{11} & A_{12}\\
A_{21} & A_{22}
\end{pmatrix}$ is an invertible 2-by-2 matrix, and $b = (b_x, b_y)^T\in \bb{R}^2$ is the translation vector. 
The prior distribution has the following form, 
\begin{align}\label{eq:oldprior}
\vect{M} \, | \, \lambda_T \sim N(\vect{M_0}, ~ \frac{1}{\lambda_T} I_2\otimes\Sigma_s^{-1}),
\end{align}
where
$M = [A ~ \mathbf{\theta}]^T = 
\begin{pmatrix}
A_{11} & A_{12} & b_x\\
A_{21} & A_{22} & b_y
\end{pmatrix}^T$, 
$M_0 = [I_2 ~ 0_{2 \times 1}]^T$ where $I_2$ is the 2-by-2 identity matrix, 
$\Sigma_s = $
$\begin{pmatrix}
\s_x^T\s_x & \s_x^T\s_y & \s_x^T\one\\
\s_x^T\s_y & \s_y^T\s_y & \s_y^T\one\\
\s_x^T\one & \s_y^T\one & \one^T\one\\
\end{pmatrix}$ where $\s_x$ and $\s_y$ consist of the $x$ and $y$ coordinates across all voxels respectively, and
$\otimes$ denotes the Kronecker product. 

The tuning parameter $\lambda_T$ can be chosen using a cross-validation approach, as proposed in \citeauthor{wang2021bayesian}, which requires intensive parallel computation. Here we propose to instead assign a hyper-prior distribution to $\lambda_T$, $\lambda_T\sim\text{Gamma}(a_T, b_T)$. 
In our work, we assign the weakly informed prior to $\lambda_T$ by choosing the hyper-parameters $a_T$ and $b_T$, in order to reflect its approximate range. Marginalizing over $\lambda_T$ yields the following prior distribution for $T_i$,
\begin{align}\label{prior:T}
    p(T_i) = \int p(T_i|\lambda_T)p(\lambda_T) d\lambda_T = t_{2a_T}\bigg(\vect{M_0}, \frac{b_T}{a_T} I_2\otimes\Sigma_s^{-1}\bigg),
\end{align}
where $t_{\nu}(\mu, \Sigma)$ denotes the multivariate t-distribution with $\nu$ degrees of freedom, location vector $\mu$, and scale matrix $\Sigma$. This fully Bayesian approach avoids the cross-validation step for determining hyper-parameters. 
Similarly we assign prior distributions for $T_i^r$ as follows:
\begin{equation}
            p(T^r_i) =  t_{2a_T^r}\bigg(\vect{M_0}, \frac{b_T^r}{a_T^r}I\otimes\Sigma_s^{-1}\bigg).
\end{equation}

The rest of parameters are assumed to follow the prior distributions:
\begin{align}
    &\alpha \sim \IG(a_{0\alpha}, b_{0\alpha}),\label{prior:alpha}\\
    &\rho\sim \text{Uniform}(L_\rho, U_\rho),\label{prior: rho}\\
    &\beta_i \, | \, \sigma^2_i\sim N(1, \lambda_0 \sigma^2_i),\label{prior: beta}\\
    &\sigma^2_i\sim \IG(a_0, a_1). \label{prior: sigma}
\end{align}

We aim to draw samples from the posterior distributions through MCMC. In particular, Gibbs sampling can be applied to the latent map $X$ and parameters $\beta_i$, $\alpha$, $\sigma^2_i$,  $\rho$, and $T_i$ are updated through Metropolis-Hasting. Details on the sampling are described in the following sections.


\subsection{Implementation of MCMC Sampling}

In this section, we describe the sampling details for the parameters of interest.

\subsubsection{Updates on the Latent Template Map with NNGP}\label{section: NNGP}
One advantage of employing the SSD as our loss function is that we are capable of deriving the closed form of the generalized full conditional distributions for $X$, $X(T)$, $\beta$, and $\sigma$ in a Gibbs sampler. Using the above model, we can derive the full conditional distribution for the subject-specific transformed latent map $X(T_i)$ for all subjects and the latent map $X$ as follows,
\begin{align*}
    &X(T_i) \, | \, \cdot \sim N(V_iU_i, V_i),\label{prior:X(T)}\\
    &X \, | \, \cdot \sim N(V_X U_X, V_X), 
\end{align*}
where $V_i= (\frac{\beta_i^2}{\sigma_i^2} I_V + S_i^{-1})^{-1}$, $U_i = \frac{\beta_i}{\sigma_i^2}Y_i + S_i^{-1}P_iX$, $V_X = (\Sigma^{-1}+\sum_{i=1}^N \big(P_i^TS_i^{-1}P_i$+$\frac{\beta_i^2}{\sigma_i^2}I_V\big))^{-1}$, and $U_X = \sum_{i=1}^N (P_i^TS_i^{-1}X(T_i) + \frac{\beta_i}{\sigma^2_i}Y_i(T_i^r))$.

Drawing samples directly from the posterior distributions defined above requires large computational burden as the number of spatial locations increase. This burden mostly arises from inverting the spatial covariance matrices $\Sigma$ and $S_i$ for  $i\in\{1,\cdots, N\}$, which requires the computational complexity of $O(V^3)$. To address this issue, the nearest-neighbor Gaussian Process (NNGP) \citep{datta2016hierarchical} provides an computationally efficient approximation to the Gaussian Process. We adapt the NNGP in the context of registration, to approximate the full conditional distributions. We describe the implementation in the following subsection. 

Here we introduce the generalization of the NNGP to group-wise registration. Let $\mathcal{S} = \{s_1, s_2, \ldots, s_V\}$ be the set of voxels indexing the latent template. We refer to $\mathcal{S}$ as the {\em template set}, and $\mathcal{T}_i = \{T_i(s_1), T_i(s_2), \ldots, T_i(s_V)\}$ as the transformed template set. We assume the sets $\mathcal{S}$ and $\mathcal{T}_i$ to be disjoint without loss of generality. This is because $Pr(\mathcal{S} \cap \mathcal{T}_i \neq \emptyset) =0$ owing to the transformation $T_i$ being continuous.
For $s_l \in \mathcal S$, let $N_{s_l}$ be the set of $m$-nearest neighbors of $s_l$ in $\{s_1,\ldots,s_{l-1}\}$,and $N_{T(s)}$ be the set of $m$-nearest neighbors of $T(s)$ in $\mathcal{S}$. 
$C(s,t)$ denotes the covariance between locations $s$ and $t$, $C(s,t)=\cov(X(s), X(t))$. $C_{N_{s_l}}$ is the covariance matrix of $X(N_{s_l})$, and $C_{s_l, N_{s_l}}$ is the covariance between $X(s_l)$ and $X(N_{s_l})$.
By the definition of the NNGP, the density of $X$ is given by: 
\begin{equation}\label{eqn: NNGP density X}
    p(X) \approx \prod_{l=1}^V N(X(s_i)|B_{s_l}X(N_{s_l}), F_{s_l}),
\end{equation}
where $B_{s_l} = C_{s_l, N_{s_l}}C_{N_{s_l}}^{-1}$ and $F_{s_l} = C(s_l, s_l)- B_{s_l}C_{N_{s_l}}B_{s_l}^T$.

The NNGP conditional density of $X(T_i)$ given $X$ and $T_i$ for subject $i$ is given as follows:
\begin{equation}\label{eqn: NNGP conditional distribution X(T) on X}
p (X(T_i)|X)  \approx \prod_{l=1}^V p(X(T_i(s_l))| X(N_{T_i(s_l)})) =\prod_{l=1}^V N(X(T_i(s_l))| B_{T_i(s_l)}X(N_{T_i(s_l)}), F_{T_i(s_l)}),
\end{equation}
where $B_{T_i(s_l)} = C_{T_i(s_l), N_{T_i(s_l)}}C_{N_{T_i(s_l)}}^{-1}$ and $F_{T_i(s_l)} = C(T_i(s_l), T_i(s_l))- B_{T_i(s_l)} C_{N_{T_i(s_l)}}B_{T_i(s_l)}^T$. In this density, the conditional distribution is decomposed as a product of the location-wise conditional densities. Each of these densities is conditional on its nearest-neighbors in $X$.

Under the NNGP model,  
the location-wise full conditional distribution of $X(T_i)$ can be derived as
\begin{equation}
    X(T_i(s_l)) \, | \, \cdot \sim N(V_{s_{il}}\mu_{s_{il}}, V_{s_{il}}), 
\end{equation}
where $V_{s_{il}} = (1/\sigma_i^2 + 1/F_{T_i(s_l)})^{-1}$, $\mu_{s_{il}} = Y_{il}/\sigma_i^2 + B_{T_i(s_l)}X(N_{T_i(s_l)})/F_{T_i(s_l)}$. We can then update $X(T_i)$ element-wise using Gibbs sampling with this full conditional distribution. As a consequence of this formulation, we can perform parallel sampling not only across subjects but also across voxels. 

Furthermore, with the NNGP densities (\ref{eqn: NNGP density X}) and (\ref{eqn: NNGP conditional distribution X(T) on X}), the conditional density of $X$ given $\{X(T_1),\ldots, X(T_N)\}$ is proportional to 
\begin{align}
p(X) \prod_{i=1}^N p (X(T_i)|X, T_i)  \approx \prod_{l=1}^V p(X(s_l)|X(N_{s_l}) \prod_{i=1}^N \prod_{l=1}^V p(X(T_i(s_l))| X(N_{T_i(s_l)})).
\end{align}

The conditional distribution of $X$ can then be derived in an element-wise manner.  For any two locations $s$ and $t$ in location space, if $s\in N(t)$ and is the $k$th component of $N(t)$, let $B_{t,s}$ be formed by the $k$th element of the row-vector $B_t$. Let $U(s_l) = \{ t \in (\cup_i \mathcal{T}_i)\cup \mathcal{S} | s_l\in N(t)\}$. For every $t\in U(s_l)$ define $a_{t,s_l} = X(t) - \sum_{s\in N(t), s\neq s_l} B_{t,s} X(s)$. Then the NNGP conditional distribution for updating $X$ is 
\begin{equation}\label{NNGP: update template}
X(s_l) \, | \, \cdot \sim N(V_{s_l}\mu_l, V_{s_l}),
\end{equation}
where
\begin{align}
    \mu_l &= B_{s_l}X(N_{s_l})/F_{s_l} +\sum_{t\in U(s_l)}  B_{t,s_l} a_{t,s_l} /F_t+\sum_{i=1}^N \frac{\beta_i}{\sigma^2_i}Y_i(T_i^r(s_l)),\\
    V_{s_l} &= (1/F_{s_l} + \sum_{t\in U(s_l)} B_{t,s_l}^2/F_t + \sum_{i=1}^N\beta^2_i/\sigma_i^2)^{-1}.
\end{align}

\noindent Updates of $X$ can be conducted sequentially with Gibbs sampling using the distribution
(\ref{NNGP: update template}). 
 
Under the nearest-neighbor Gaussian Process described above, the complexity of computing the posterior distributions reduces to $O(Vm^3)$. As in \cite{datta2016hierarchical}, we can get a reasonable approximation to the full likelihood using a neighborhood size of $m\geq 10$, and it provides a significant computational improvement compared to $O(V^3)$ using the full model. 

\subsubsection{Dynamic Neighborhood Updates}
The standard NNGP model was designed using a fixed design, i.e., a fixed collection of spatial locations, and the neighborhood membership set was constructed prior to performing sampling using MCMC. \citet{dnngp} proposed dynamically updating the neighborhoods in the context of spatio-temporal data, but the algorithm benefited from having a fixed set of spatial locations. In the registration problem,  the set of locations $T_i(\mathcal{S})$ vary in each iteration of MCMC based on the current value of the transformation $T_i$. As a consequence, the standard NNGP cannot be applied seamlessly to the registration problem, as the updated transformations lead to dynamically updated spatial locations for the transformed template. Therefore, the neighbor set of $\mcl{T}_i$ in $\mcl{S}$ should be derived when the transformation is updated, which requires $O(V)$ flops \citep{Finley2019} for each $T_i(s_l)$, $i=1,\ldots,N$, $l=1,\ldots,V$ and will increase the computational complexity of sampling. 

We propose an efficient approach towards finding the nearest neighbors within each iteration leveraging the lattice structure of the voxels. The idea is to create a library of the neighbors for each possible location at the initialization stage, from which the neighbors of the transformed locations can be easily looked up. It consists of three steps. First, we generate an enlarged grid, $L(\mcl{S})$, by adding more spatial locations around the grid $\mcl{S}$. The choice of $L(\mcl{S})$ is such that all reasonable transformations $T(\mathcal S)$ would lie within the boundary of the grid  $L(\mcl{S})$. Second, we find the $m$-nearest neighbors in $\mcl{S}$ for each element of $L(\mcl{S})$. This is a one-time task and these neighbor sets are pre-computed before running the MCMC. Finally, during each iteration of the MCMC, for the current value of each location in $\mcl{T}_i$, we find the closest point in $L(\mcl{S})$, and extract the neighborhood information. Notice as we are working with images, every point is located on a regular grid. Hence, we can find the index of the closest point of any $s$ in $L(\mcl{S})$ without explicitly computing pairwise distances between $\mcl{T}_i$ and  $\mcl{S}$. For example, assuming the grid points have integer-valued coordinates, we can get the coordinates of the closest grid point of a transformed location by rounding to the nearest integer, along with its index in $L(\mcl{S})$.  

\begin{figure}
    \centering
    \includegraphics[width=.8\textwidth]{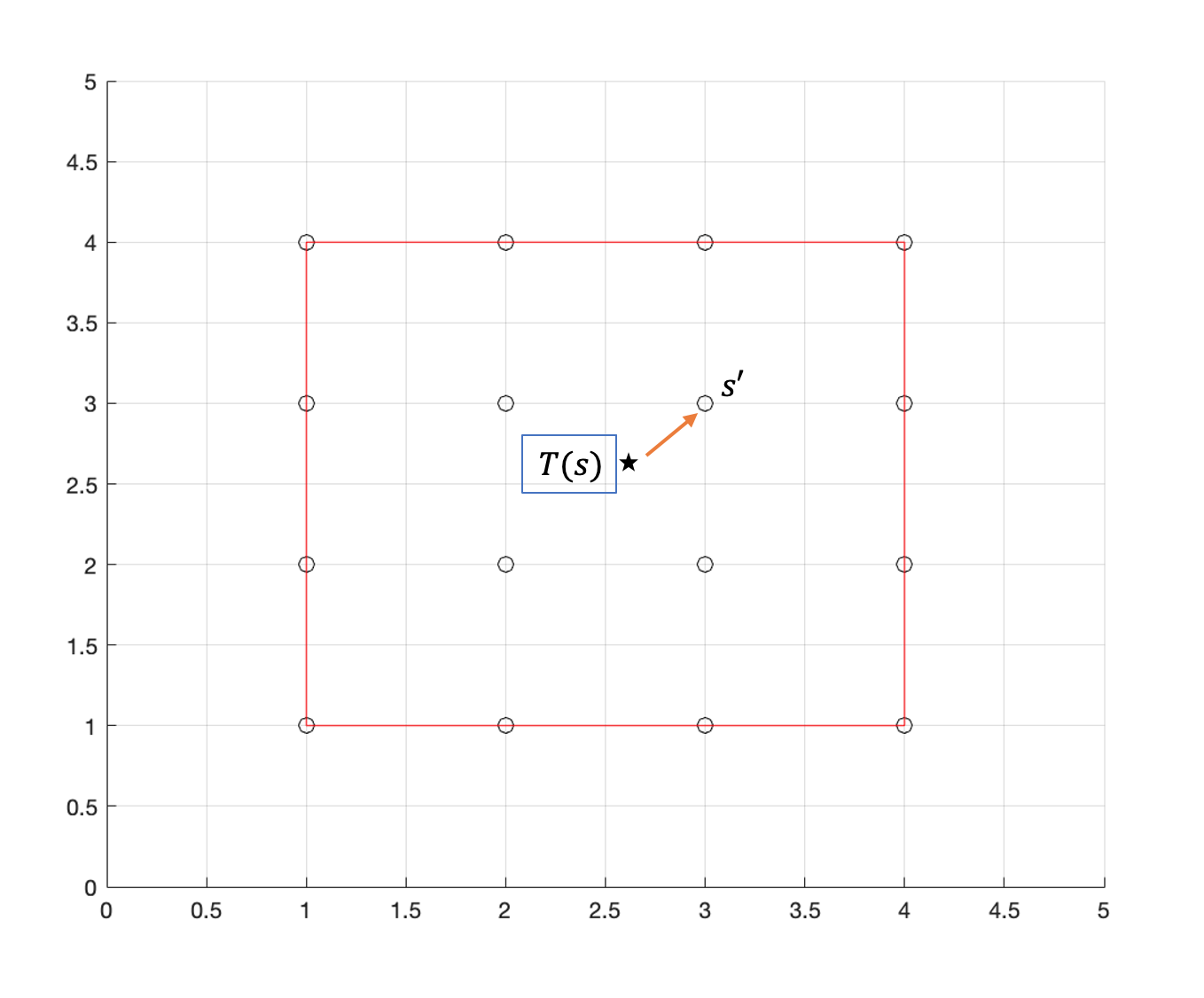}
    \caption{Illustration of the proposed dynamic neighborhood search.  The enlarged template set, $L(\mathcal{S})$, is denoted by locations (black circles) on the regular grid. The neighborhoods for each location in $L(\mathcal{S})$ are precomputed. We approximate the neighborhood of the transformed location using the neighborhood of its closest location on $L(\mathcal{S})$. In this example, the neighborhood set of $T(s)$ is approximated by $N_{s'}$.}
    \label{fig:dynamicNNGP}
\end{figure}

\pink{Figure \ref{fig:dynamicNNGP} shows a schematic visualization of the dynamic neighborhood search in the two-dimensional scenario. The enlarged template set includes locations on the regular grid, denoted by the black circles. As an example in the figure with an integer-value grid, the transformed location $T(s)=(2.6, 2.6)$ has the closest point $s'=(3,3)$. Then the nearest neighbors of $T(s)$ is approximated by $N_{s'}$, which has been collected in the beginning. 
}

While the resultant neighbors $N_{T(s)}$ will not be the exact $m$-nearest neighbors of $T(s)$ in $\mathcal S$, they are guaranteed to be proximal to $T(s)$ as they are the $m$-nearest neighbors of the point in $L(\mcl{S})$ that is closest to $T(s)$. Only the proximality of the neighbor sets are important to the success of NNGP as an approximation and exact nearest neighbors are not a requirement. Hence, this pragmatic dynamic neighbor search saves a significant amount of computational effort without substantial compromise.

\subsubsection{Updates on the Transformations $T_i$}\label{section: update T_i}

We follow the Metropolis-Hasting algorithm for sampling the transformations $T_i$. Using the NNGP, the likelihood involving $T_i$ can be approximated as follows:
\begin{align}
\label{eqn: likelihood of T}
p(T_i) p(X(T_i)|X, T_i)  &\approx p(T_i)\prod_{l=1}^V p(X(T_i(s_l))| X(N_{T_i(s_l)}))]\nonumber\\
& =p(T_i)\prod_{l=1}^V N(X(T_i(s_l))| B_{T_i(s_l)}X(N_{T_i(s_l)}), F_{T_i(s_l)}),
\end{align}
where $p(T_i)$ denotes the prior distribution of $T_i$, defined in section \ref{section: prior distributions}.

\pink{The space of transformation T is generally not linear. For example, addition of two affine transformations does not results in another affine transformation. For this reason, the proposal distribution for the transformation $T$ is updated on the corresponding tangent space where it is feasible to apply linear operations. For another familiar instance, assuming the parameters of interest are restricted to be positive, one reasonable proposal is performed on the logarithm of the parameters. Here the logarithm function maps the parameter onto its tangent space. 
}

For the affine transformation, following \cite{Eade2017}, we can sample the transformation at the $k^{th}$ iteration, $T^k$, from the proposal distribution with mean $T^{k-1}$ and covariance matrix $\lambda_\delta\Sigma_T$, where $\lambda_\delta\in \bb{R}^{+}$ controls the step-size and $\Sigma_T\in\bb{R}^{6\times 6}$, by performing two steps. First, we draw a sample $\delta\in\bb{R}^6$ from the multivariate normal distribution with mean $0$ and variance-covariance matrix $\lambda_\delta\Sigma_T$, i.e., $\delta\sim N(0, \lambda_\delta\Sigma_T)$. Second, we  get the proposal through the composition of $\exp(\delta)$ and $T^{k-1}$, i.e., $T^k = \exp(\delta)\cdot T^{k-1}$, where $\exp(\delta)$ is the exponential mapping of the Lie algebra $\delta$. In our case, $\delta$ can be derived by the matrix logarithm of the 3-by-3 affine transformation matrix,
$\begin{pmatrix}
A & b\\
0 & 1
\end{pmatrix}$, and $\exp(\delta)$ is the inverse mapping.

The sampling steps \pink{on the tangent space fundamentally }include a change of variable. To handle this, we follow the results of \cite{falorsi2019reparameterizing} who formulated the probability density of $T^k$ given $T^{k-1}$ to be:
\begin{align}
p(T^k|T^{k-1}) = p(\delta)|J(\delta)|^{-1},\;\; \text{with}\;\; J(\delta) = \prod_{\substack{\lambda\in Sp(\text{ad}_\delta),\\ \lambda\neq 0}}\frac{\lambda}{1-e^{-\lambda}},
\end{align}
where $Sp(\text{ad}_\delta)$ denotes the eigenvalue spectrum of the adjoint representation of $\delta$, and $p(\delta) = N(0, \lambda_\delta\Sigma_T)$.
To further improve the mixing time, we applied the adaptive Metropolis algorithm \citep{shaby2010exploring}, to manipulate both $\lambda_\delta$ and $\Sigma_T$ adaptively. The details are referred to the original work and omitted here.

In order to ensure the identifiability of the latent map and the transformations, the updates of $T_i$ at each iteration are standardized to the midpoint space \citep{Arsigny2006, balci2007free}.
The standardization for the transformation on each subject can be derived by the composition of the mean map, 
\begin{equation}
    \hat T_i = T_i \circ \bar T^{-1},
\end{equation}
where $\bar T$ is the mean map of the transformations ${T_1,T_2,\ldots, T_N}$, which can be derived using the iterative procedure described by \cite{fletcher2003gaussian}. Notice that the standardization step reduces the degree of freedom. It is equivalent to sampling the initial $N-1$ transformations only, because the last transformation will be fully determined by the first $N-1$ transformations.


The scaling-correction parameters $\beta_i$, subject-specific noise terms $\sigma_i^2$, and spatial variance $\alpha$ are routinely updated through Gibbs sampling. The spatial decay parameter $\rho$ is updated through the random-walk Metropolis steps. Details can be found in the supplemental materials.

\section{Simulation}\label{Section: Simulation}

Our framework can be applied to multi-dimensional registration problems. In this section we will perform simulations in both one-dimensional and two-dimensional settings, to show the effectiveness of our proposed method. 

\subsection{One-dimensional Curve Registration}
\begin{figure}
    \centering
    \includegraphics[width =\textwidth]{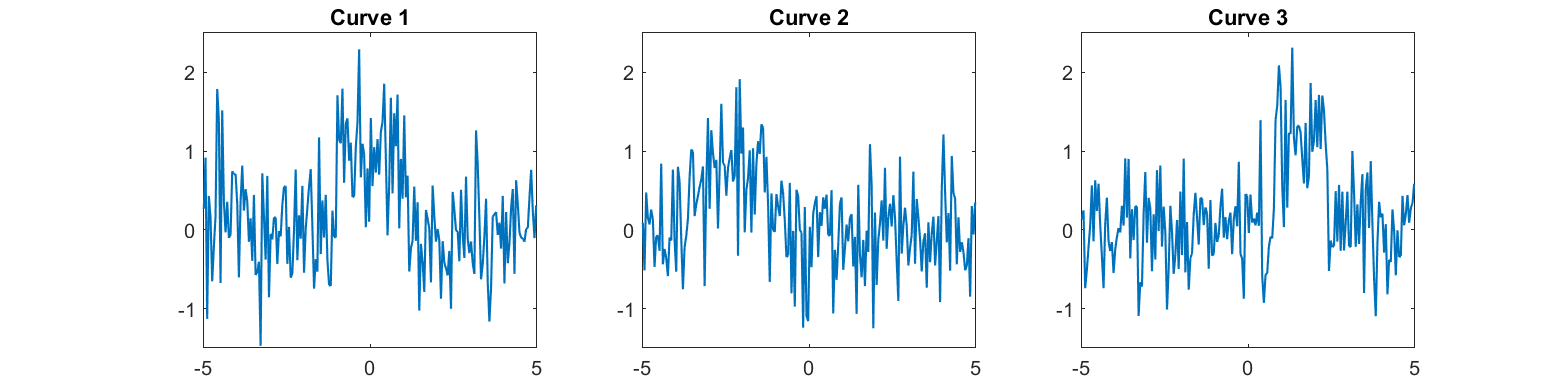}
    \caption{Three raw curves generated from the indicator function, $X(s)=I_{[-1\leq s\leq 1]}$, by shifting and scaling the template and adding IID noise with variance $\sigma^2 = 0.5^2$. }
    \label{fig:raw_indicator}
\end{figure}

Let the template $X$ be the indicator function on the interval $[-5,5]$, $X(s)=I_{[-1\leq s\leq 1]}$. Three curves, shown in Figure $\ref{fig:raw_indicator}$, are generated by shifting and scaling the template and adding IID noise with variance $\sigma^2 = 0.5^2$. The curves are sampled on an equally-spaced grid with spacing $0.05$. For the conventional model of \cite{allassonniere2007towards}, we select the landmarks at locations $0.1$ apart, and place Gaussian kernels with width $0.1$. For our probabilistic model, we choose the NNGP neighborhood size to be $10$. 
The prior distributions are assumed to be $\alpha\sim\IG(0.2, 0.1)$, $\rho\sim\text{Unif}(0, 3)$, $\beta_i|\sigma^2_i\sim N(1, \sigma^2_i)$ and $\sigma^2_i \sim \IG(2,1)$. The hyperparameters defining the prior distribution of the transformations are set to be $a_T = 0.1$ and $b_T = 0.1$.

\begin{figure}
    \centering
    \includegraphics[width =.4\textwidth]{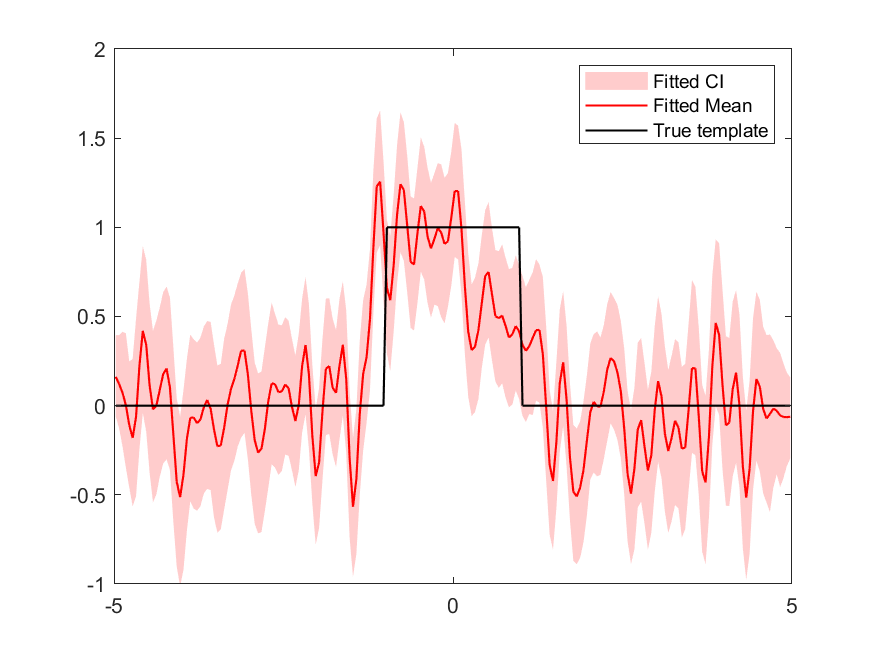}
    \includegraphics[width =.4\textwidth]{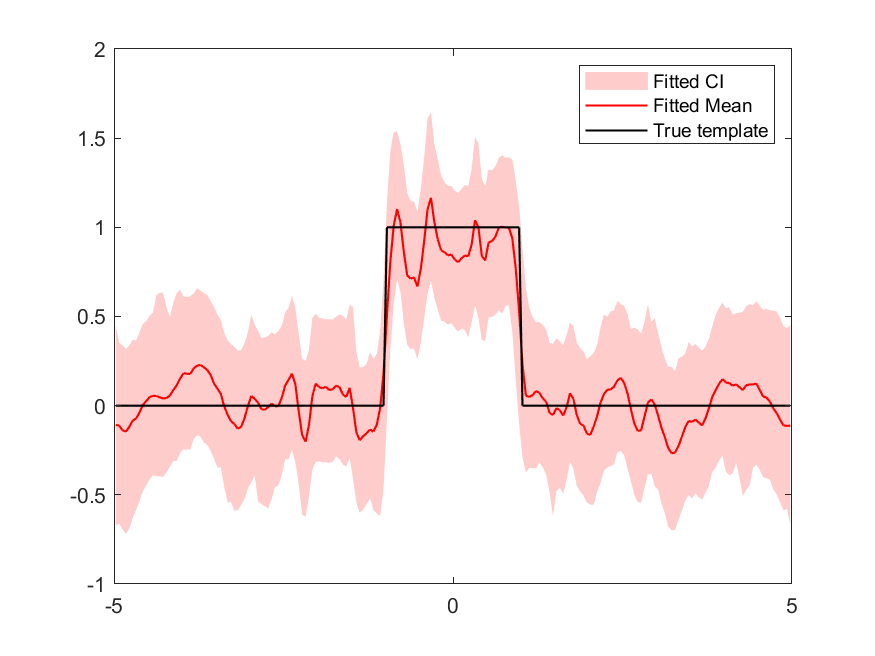}
    \caption{Comparison between the conventional probabilistic method (left) and the proposed method (right). The truth (in black) is overlaid on both figures. The conventional method performed worse in the area around the edges of the signal. The same phenomena was discussed in \cite{allassonniere2007towards}. The proposed method succeeds in recovering the truth, which is covered by the credible intervals.}
    \label{fig:simulation_1d_indicator}
\end{figure}

Figure $\ref{fig:simulation_1d_indicator}$ shows the posterior means (red) and the point-wise credible intervals (pink) for the conventional model described in Section \ref{Section: Review} and our proposed model. The true template is also drawn for reference. The conventional approach does poorly in the template estimation near the boundaries between the two intensity levels. This phenomenon has been described in the original work. The results on the right of the figure shows a better performance using our proposed approach.

\begin{figure}
    \centering
    \includegraphics[width =\textwidth]{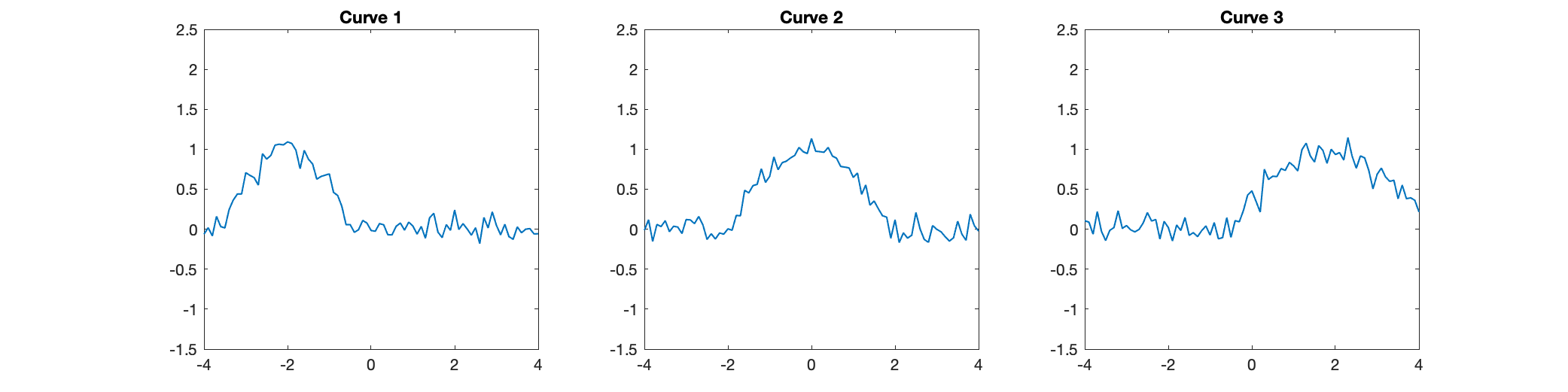}
    \caption{Three raw curves generated from the template curve, which is a truncated cosine function defined by $X(s) = I(|s|<L)\cos{(\frac{\pi s}{2L}})$, for $L>0$. A set of curves were generated by shifting and scaling the template, $X(\beta_0+\beta_1 s)$, and adding IID noise with variance $\sigma^2 = 0.1^2$.}
    \label{fig:raw_cosine}
\end{figure}

In the second example, we compare the methods in the case of a smooth template and lower noise levels. Let the template be the truncated cosine curve defined by $X(s) = I(|s|<L)\cos{(\frac{\pi s}{2L}})$, for $L>0$. We generate a set of curves by shifting and scaling the template, $X(\beta_0+\beta_1 s)$, and adding IID noise $\epsilon(s) \sim N(0, 0.1^2)$. The curves are generated on the domain $[-4, 4]$, sampling on the grid with $0.1$ units separating points. Figure \ref{fig:raw_cosine} shows the three generated curves. For the conventional method, the landmarks are chosen at each grid and the Radial Basis Kernel has the smoothing parameter equal to $0.05$. We choose the same set of prior distributions as the first scenario for the proposed probabilistic method.

\begin{figure}
    \centering
    \includegraphics[width =.4\textwidth]{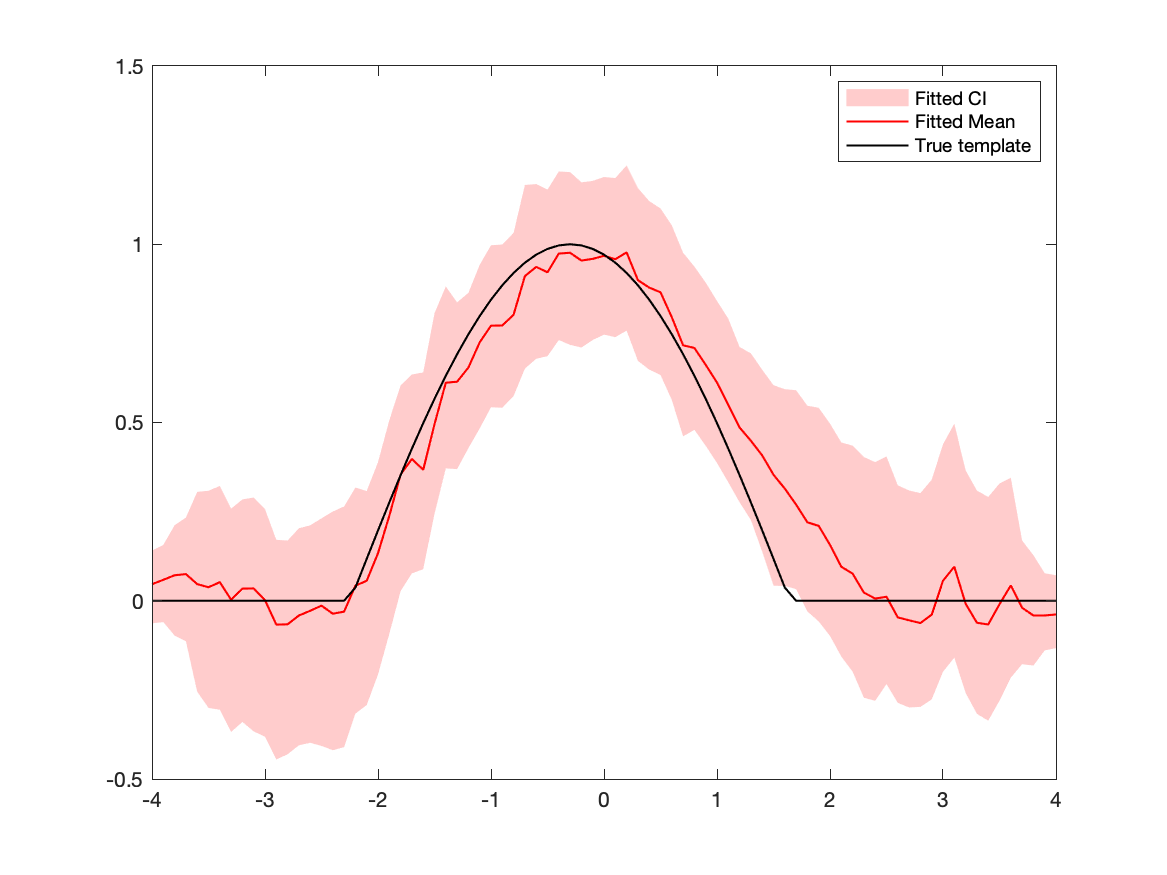}
    \includegraphics[width =.4\textwidth]{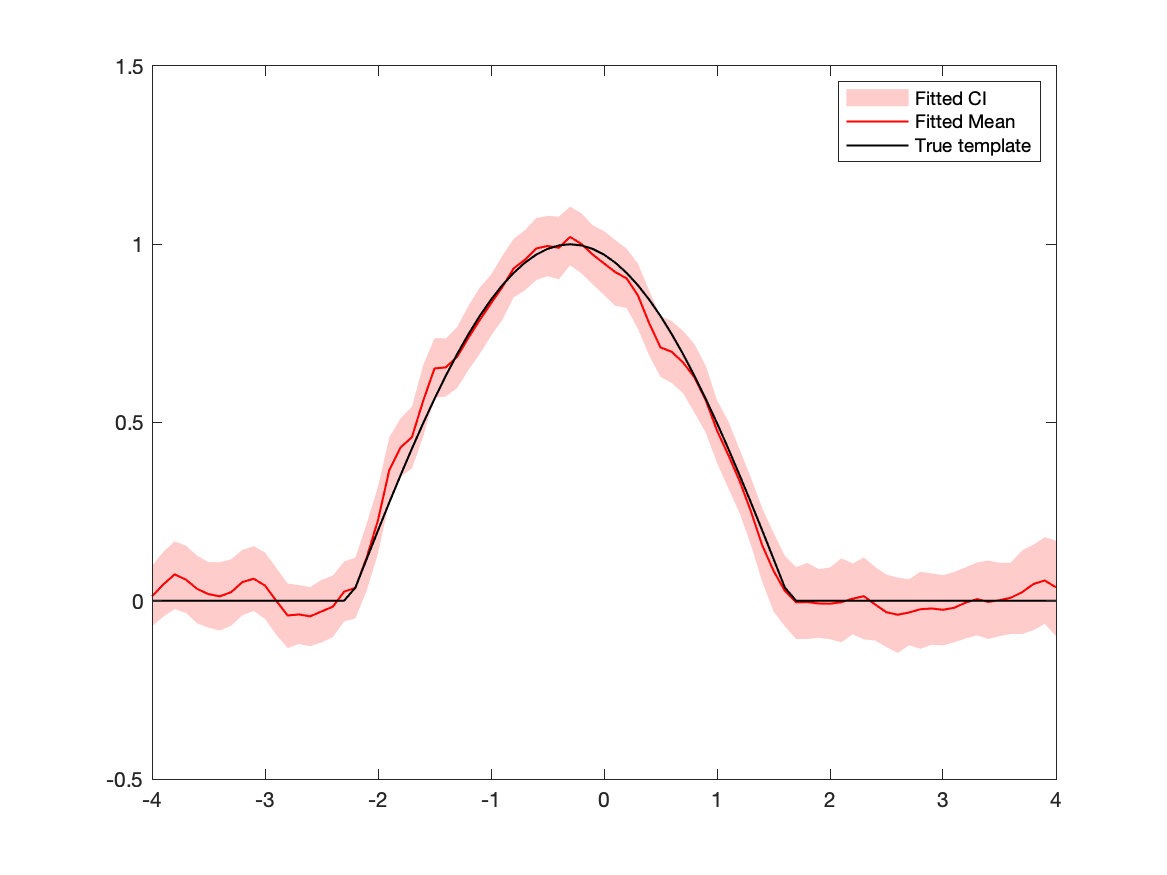}
    \caption{Comparison between the conventional method (left) and the proposed method (right). The truth (in black) is overlaid on both figures. The conventional method produced wider confidence intervals and performed worse in the area around the edges of the signal than our proposed probabilistic method.}
    \label{fig:simulation_1d_consine}
\end{figure}

In Figure \ref{fig:simulation_1d_consine}, we show the results of both the conventional method and our proposed method. Sample median are plotted along with the location-wise credible interval. Both methods correctly recover the true template. The conventional method produces a template estimation with high heteroscedasticity across locations, 
and performs poorly near the boundary (i.e., $s=2$). However, the results obtained using our method shows a smoother and narrower credible band, which is the consequence of the Gaussian Process model that better captures the spatial pattern. \pink{The tight credible band also indicates an improvement in statistical sensitivity for detecting the template.}  

\subsection{Two-dimensional Image Registration}
\begin{figure}
    \centering
    \includegraphics[width=\textwidth]{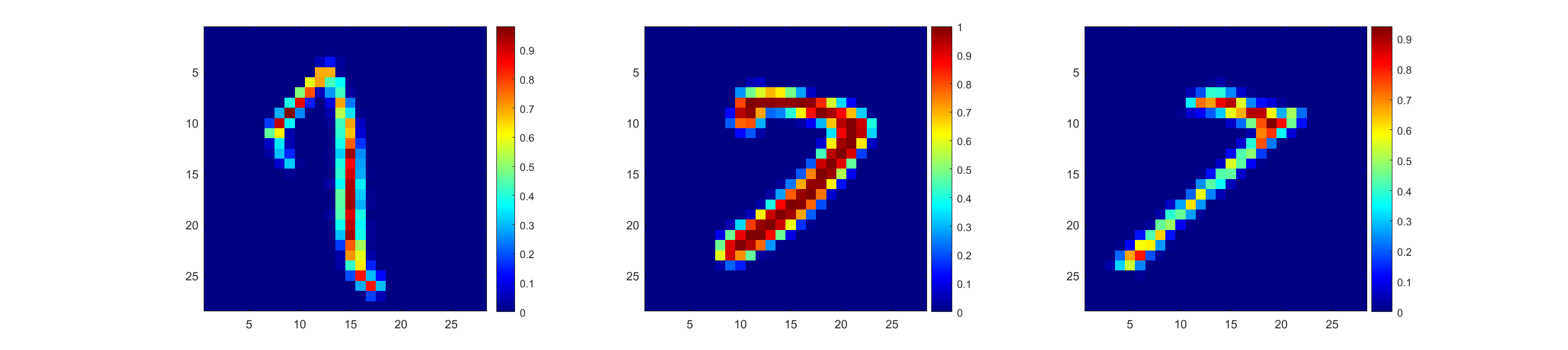}
    \caption{Raw images used for the simulation in the two-dimensional case.}
    \label{fig: 2d raw imags}
\end{figure}

We further illustrate the effectiveness of our proposed method on two-dimensional images through an application to images from the Modified National Institute of Standards and Technology (MNIST) database. The MNIST database is a large database of handwritten digits that is commonly used for training various image processing systems \citep{lecun1998gradient}. The processed data set can be found in the Deep Learning Toolbox in \cite{MATLAB:2021a}. Each image has 28-by-28 pixels and has been rotated by a certain angle. Here we choose three images of ``7" as a proof-of-concept, as shown in Figure \ref{fig: 2d raw imags}. 

For the conventional probabilistic approach, we select the landmarks on a regular two-dimensional grid which is twice coarser than the image domain, to make the computational effort comparable. The width of Gaussian kernel is set to be $1.5$. For our proposed probabilistic approach, we chose the neighborhood size to be $10$. The prior distributions are assumed to be $\alpha\sim\IG(2, 1)$,  $\rho\sim\text{Unif}(0, 3)$, $\beta_i|\sigma^2_i\sim N(1, \sigma^2_i)$ and $\sigma^2_i \sim \IG(2,1)$. The hyperparameters defining the prior distribution of the transformations are set to $a_T = 2$ and $b_T = 1$. The estimated latent map was assumed to have the large domain with the extension of $5$ pixels to each of the boundaries, so that we are able to capture the excessive variability outside of the original domain. 

\begin{figure}
    \centering
    \includegraphics[width=\textwidth]{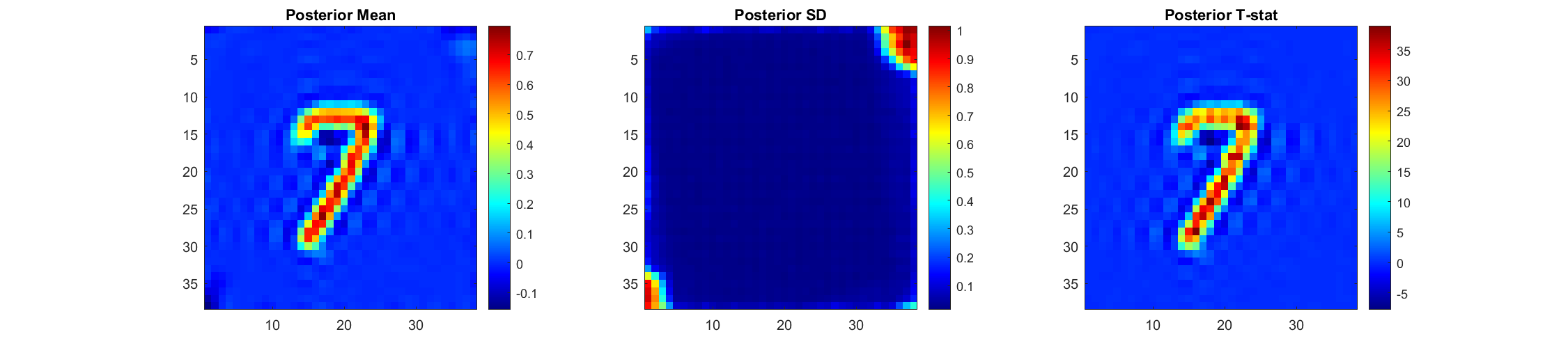}
        \includegraphics[width=\textwidth]{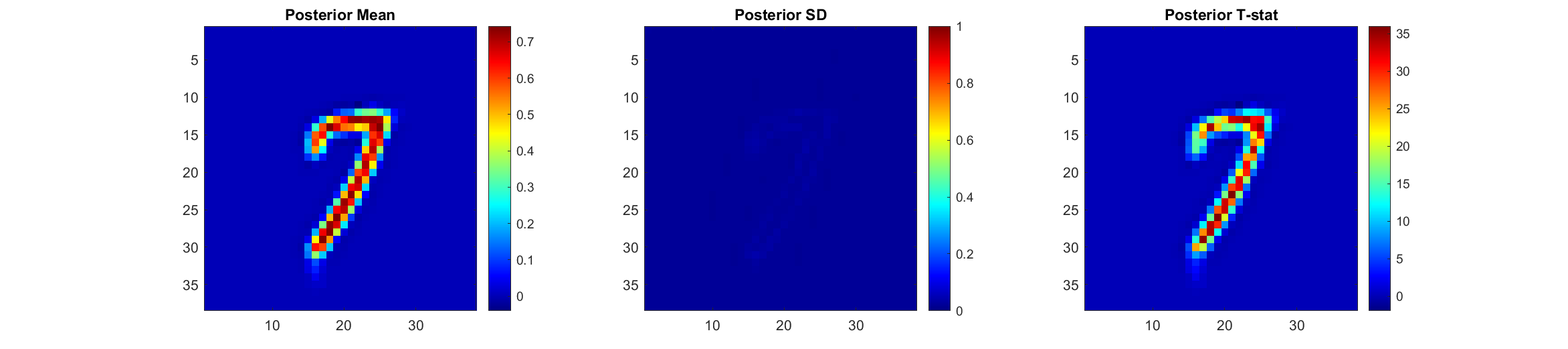}
    \caption{Results from the conventional probabilistic method (top row) and the symmetric registration model (bottom row). The results obtained using the conventional method show  patterns of fluctuations in the background, and the variation explodes in the corner regions. On the other hand, the proposed method shows more reasonable results for the posterior mean (left column). In the plot of the posterior standard deviations (middle column), the proposed approach shows less variability.}
    \label{fig: 2d simulation results}
\end{figure}

Figure \ref{fig: 2d simulation results} shows the results obtained using both the conventional and the proposed approach. The results consist of comparisons for the posterior mean (left column), pixel-wise standard deviation (middle column) and the ratio between them (right column). The conventional method produces bias in the background of the template estimation as manifested in the distinct chequered patterns. This is the result of the linear combination of the radial Gaussian functions. In contrast, our proposed method shows a much cleaner template estimation. The standard deviation blows up at the corners for the conventional method, which indicates areas of high uncertainty. For our proposed approach, the standard deviation is lower at the center of the map where more information is stacked, and is higher at the corner. This is due to the affine transformation of the original images. We find that a counter-clockwise rotation is performed on the template to be matched with the first image, whereas clockwise rotation is required for the other two images. This process makes the template less informative and more uncertain in the corner area. This spatial variability in the uncertainty is thus expected and desirable and appears in the results of the proposed method, but not the conventional approach.

\begin{figure}
    \centering
    \includegraphics[width=\textwidth]{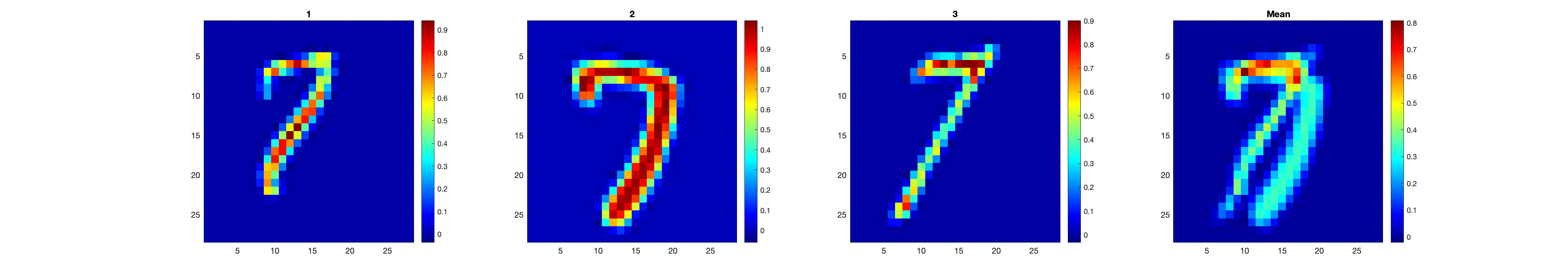}
        \includegraphics[width=\textwidth]{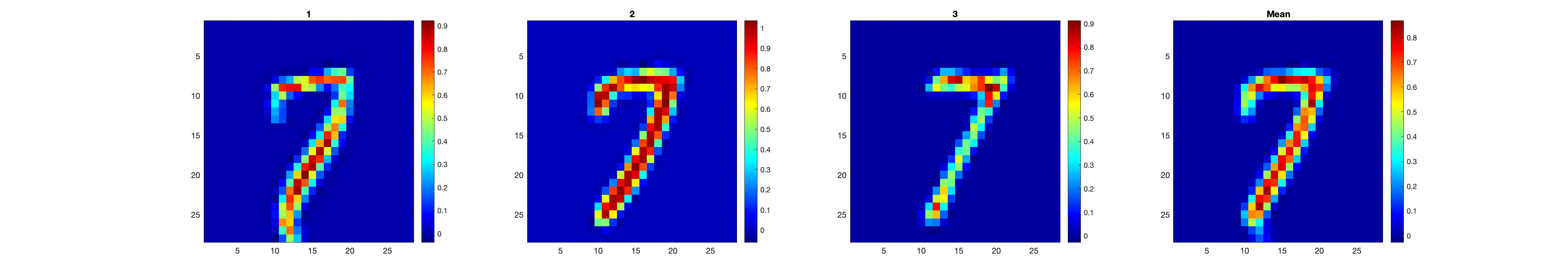}
    \caption{Comparison between the conventional model (top row) with the symmetric registration model (bottom row). Inversely deformed images are illustrated in the first three columns. The fourth column shows the average image. Clearly, the registered images from the probabilistic approach are misaligned in contrast to the symmetric regression approach.}
    \label{fig: 2d simulation inverse results}
\end{figure}

We also transformed the subject-specific images with the inverse of the mean of transformation from the probabilistic model.
Figure \ref{fig: 2d simulation inverse results} compares the inversely transformed images obtained using the probabilistic model (top row) with images obtained using the symmetric registration model. The first three images are the individual maps transformed by the inverse transformation, and the last image in each row is the mean across the transformed images. We can see that for the conventional method (top-row) the subject-specific images are not aligned properly with the inverse of the forward registration. The images inversely transformed by our inverse-consistent symmetric registration framework does not have this issue and are well aligned which in turn improve subsequent group-level inference based on these transformed images.

\section{Application to Data from the Thermal Pain Study}\label{Section: Application to the pain data}

\green{Functional group-wise registration using our proposed model was applied to the thermal pain data described in Section \ref{Section: Pain data description}.} The symmetric group-wise registration model was fit using the generalized Bayes framework to get the posterior samples for both the latent map $X$ and the transformations $T_i$ and $T_i^r$ for all subjects. The prior distributions were assumed to be $\alpha \sim \IG(2, 1)$,  $\rho \sim \text{Unif}(0, 3)$, $\beta_i|\sigma^2_i \sim N(1, \sigma^2_i)$ and $\sigma^2_i \sim \IG(2,1)$. The hyperparameters defining the prior distribution of the transformations were set to $a_T=a_T^r = 0.1$ and $b_T = b_T^r = 0.1$. The spatial domain of the latent map was extended by $5$ voxels at each edge in order to allow for the identification of uncertainty outside the original domain of maps due to excessive transformation. MCMC sampling was run for $3\times 10^4$ iterations and the initial $1.5 \times 10^4$ iterations were disregarded in the burn-in stage.

Figure \ref{fig: real, fitted latent map} summarizes the posterior distributions of the latent map. The left panel shows the voxel-wise mean map, which demonstrates three islands with different activation levels. In the middle panel, where we plot the standard deviation of the posterior latent map, there exists a spike at the location of the right islands, which provides evidence of its high uncertainty. The three isolated peaks remain sharp in the ratio plot shown in the right panel. Figure \ref{fig: real, inversely warped maps} exhibits the transformed subject-specific mean activation maps. 

\green{Using both the standard unregistered maps and the reversely transformed activation maps, two analyses were performed to evaluate the benefits of functional registration. First, we performed a group-level t-test to determine whether there exists a significant population-wide activation related to the painful stimuli. This corresponds to the analysis performed in a standard brain mapping study.  Second, we trained a model to predict reported pain. This corresponds to the 
analysis performed in a predictive modeling approach, and is the main goal of this work. The results of these analyses are described in the following sections.}

\subsection{Brain Mapping}

\green{We first performed a group-level fMRI analyses concerned with determining whether there exists significant population-wide activation related to the painful stimuli.  Standard group analyses typically involve fitting two separate models. A first-level GLM analysis is performed on each participant’s data, providing within-subject functional activation maps. A second-level analysis provides population inference on whether there are non-zero effects \citep{lindquist2012estimating}. Researchers typically perform this analysis using a t-test across participants at each voxel of the brain. These types of analyses are critical for brain mapping studies which help describe the local encoding of information in the brain.}

\begin{figure}
    \centering
    
    \begin{subfigure}[ht]{.8\textwidth}
    \includegraphics[width=\textwidth]{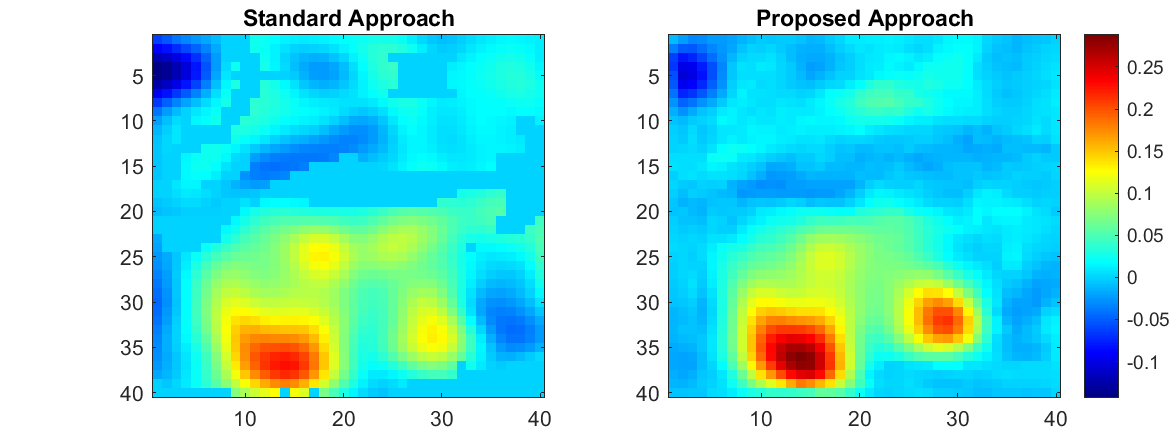}
    \caption{Mean maps}
    \label{fig:summary plots comparisons: mean}
    \end{subfigure}
    
    \begin{subfigure}[ht]{.8\textwidth}
    \includegraphics[width=\textwidth]{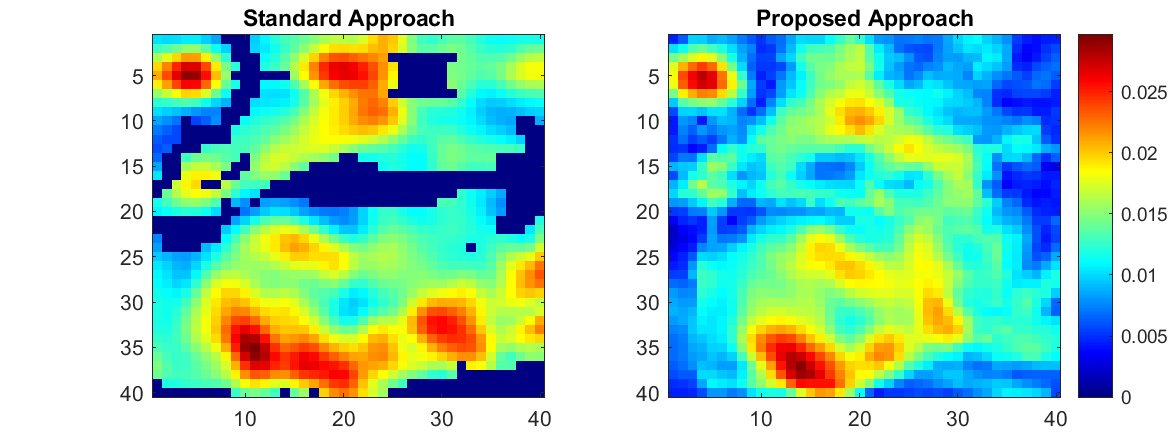}
    \caption{Standard deviation maps}
    \label{fig:summary plots comparisons: sd}
    \end{subfigure}
    
    \begin{subfigure}[ht]{.8\textwidth}
    \includegraphics[width=\textwidth]{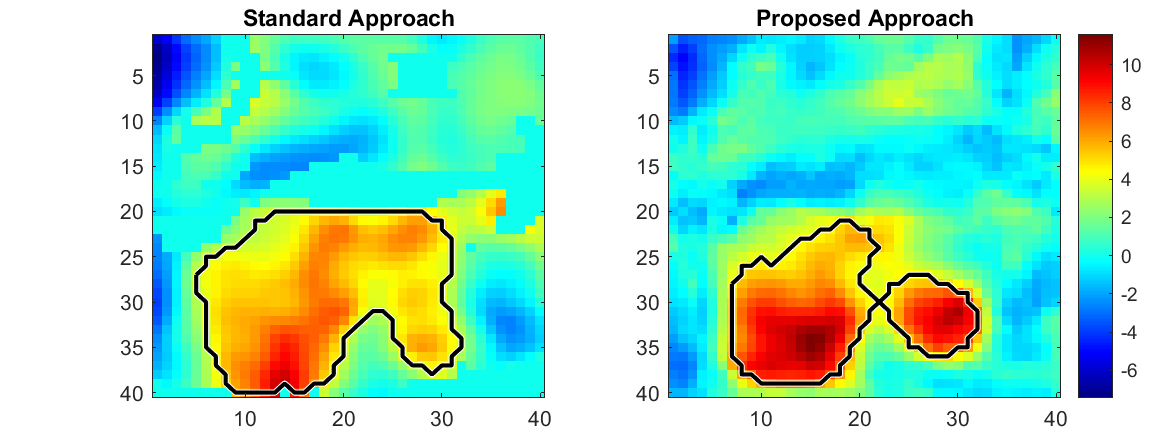}
    \caption{T-statistic maps}
    \label{fig:summary plots comparisons: tstat}
    \end{subfigure}
    
    \caption{Comparison of group-level inference, including (a) mean maps, (b) standard deviation maps, (c) t-statistic maps. We compare the standard unregistered approach (left) with the proposed method (right). In the t-statistic maps the most activated regions are enclosed within the black lines.}
    \label{fig:summary plots comparisons}
\end{figure}

\green{We compare the results of the second-level analysis using the standard and transformed activation maps. Figure \ref{fig:summary plots comparisons} shows maps of the mean, standard deviation, and t-statistic values. In Figure \ref{fig:summary plots comparisons: mean}, we find that the transformed activation maps provide higher average values within brain regions expected to be activated. The standard deviation maps in Figure \ref{fig:summary plots comparisons: sd} show higher variability in larger areas of the brain under the conventional analyses. Figure \ref{fig:summary plots comparisons: tstat} compares the t-statistic maps, indicating higher sensitivity for spatial localization using the proposed approach. Segmentation techniques, for example the Gaussian mixture model, can be used to detect the activation regions. As shown in Figure \ref{fig:summary plots comparisons: tstat}, the detected activation areas are highlighted in black. We found that the t-statistic map obtained using the proposed method produced isolated activation areas that conform to known somatosensory areas, while the conventional approach failed to do so.}

\subsection{Pain Prediction}

\green{Next, a model was trained to predict pain intensity. }
In previous work, machine-learning models have been developed to predict pain intensity using the functional activation maps as covariates \citep{wager2013fmri, lindquist2017group}. For example, \citet{wager2013fmri} proposed a Lasso-PCR model, which integrates the principal components regression model with an $L^1$-penalty on the coefficients. Leave-one-subject-out cross-validation steps was utilized to predict pain scores for each subject. They also evaluated the intra-subject correlations between the predicted and true pain scores. 

\pink{To evaluate the effectiveness of our proposed approach in the context of prediction, we fit a similar prediction model using the registered functional activation maps. 
We utilized a leave-one-subject-out cross-validation procedure, which consisted of the following steps.
First, a training set was constructed using all but one subject's activation maps, and the left-out subject's maps formed the testing set.
Second, using the training set, we fit our proposed model to generate posterior samples of the latent template and the subject-specific transformations. The transformed maps in the training set can be derived by transformation with the posterior means. We also found the estimation of the latent template using its posterior mean. 
Third, transformed functional activation maps were created for the test set by registering the test maps to the latent template. Fourth, we trained the Lasso-PCR model on the transformed maps in the training set and made predictions using the transformed maps in the testing set. By iterating over all subjects, we got predictions of the pain intensity corresponding to each trial and subject.
}

Figure \ref{fig: real, correlation plots, violen plot} compares the correlation between predicted and actual pain ratings obtained using the models fit on the standard activation maps and the transformed maps. The correlations for the original activation maps are more distributed, and some even take negative values. The mean correlation obtained using the standard activation maps is $0.5796$, while the prediction model trained using the transformed maps produced a correlation of $0.7070$ between the predicted and true pain scores. This demonstrates a significant improvement in prediction accuracy when using the transformed activation maps. 

Figure \ref{fig: real, correlation plots, scatter plot} takes a closer look at the comparison. Subjects were clustered based on the prediction-outcome correlations obtained using the model trained on the standard activation maps. A threshold of $r=0.5$ was used to define clusters. We found that $81.82\%$ of the subjects in the group where the predictions using the standard maps produced poor results ($r<0.5$) showed an improvement when using the transformed images. In the other group, where predictions using the standard images produced stronger correlations with the true pain scores, $50\%$ of subjects showed further improvement when using the transformed images. The results indicate that the functional registration significantly improved the performance of a prediction model in outlier cases, 
while they still produce comparable predictions for cases the are more well-behaved.

\section{Discussion}\label{Section: Discussion}
In this work, we propose a Bayesian framework for reducing the inter-subject misalignment in functional topology in the brain. Using inverse-consistency, our approach is capable of simultaneously estimating the latent activation template map and the registration functions for each subject. We model the activation maps with a spatial Gaussian process that explicitly models the shape and extent of fMRI activation. The full Bayesian formulation of the model enables inference on the latent surface via posterior samples. Our generalized Bayes framework extends the probabilistic registration approach to achieve symmetric registration, allowing us to obtain transformed subject-specific activation maps. We generalize the nearest-neighbor Gaussian process model to the group-wise registration framework, in order to accomplish computational efficiency while dynamically updating transformed voxels and their neighbor sets. In an application to fMRI data from a study of thermal pain, we found that the registration procedure contributes to increased sensitivity for group-level inference and improves population-level models of the functional brain representations for predicting pain score ratings. 

There are some limitations in the current implementation of our approach. First, as the proposed approach is designed for registration within a local region of interest, the suitability of the approach for multi-region registration needs to be studied.
Moreover, our proposed approach when using affine transformations may suffer in the case that there exists significant non-linear discrepancy between the subject-specific activation maps. However, the methodology is general and can work with any more general class of non-linear transformations.

For the purpose of simplifying the description of our approach, we illustrated our approach using a two-dimensional version in this paper. However, a three-dimensional version can be constructed analogously. In practice, we would recommend using a 3D implementation when 3D data is available.

In future work, as our data experiment measures multiple brain maps at six different temperatures, we will model the within-subject variability using the registration and explore its association with other effects such as temperature. This will allow us to explore how the spatial extent of brain activation changes as a function of stimulus intensity. This provides an avenue to move past simply looking at mean spatial intensity in a specific voxel as a measure of brain activation. Another possible extension is evaluating the multilevel variability through inference on the within-subject and between-subject transformation.

\section*{Acknowledgments}
The work presented in this paper was supported in part by NIH grants R01 EB016061 and R01 EB026549 from the National Institute of Biomedical Imaging and Bioengineering
and R01 MH116026 from the National Institute of Mental Health, and National Science Foundation Division of Mathematical Sciences grant DMS-1915803. 

\newpage
\bibliographystyle{apalike}
\bibliography{reference}

\newpage



\begin{figure}
    \centering
    \includegraphics[width=\textwidth]{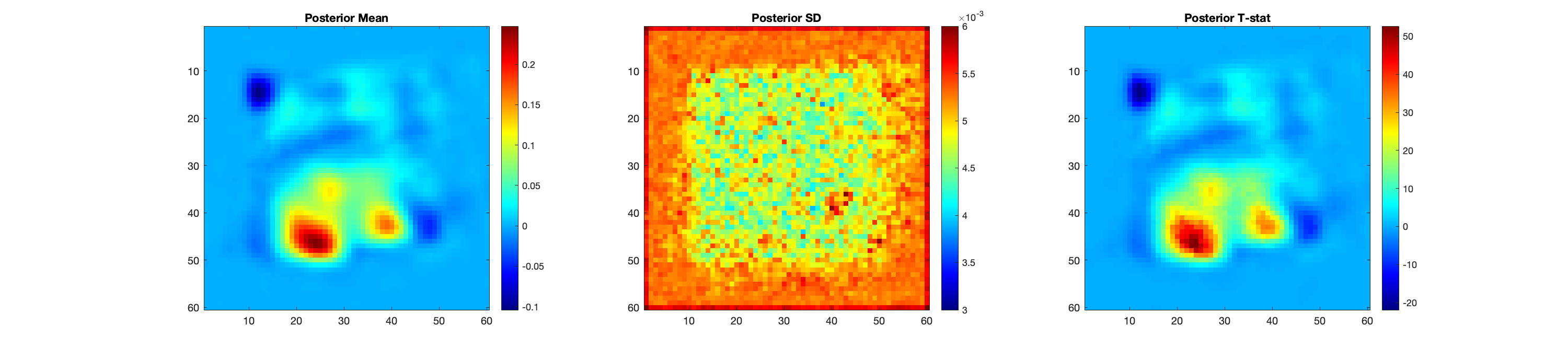}
    \caption{Summary plots of the posterior samples of the latent template map: Mean (left), Standard Deviation(middle), Ratio of the mean over standard deviation (right).}
    \label{fig: real, fitted latent map}
\end{figure}

\begin{figure}
    \centering
    \includegraphics[width=\textwidth]{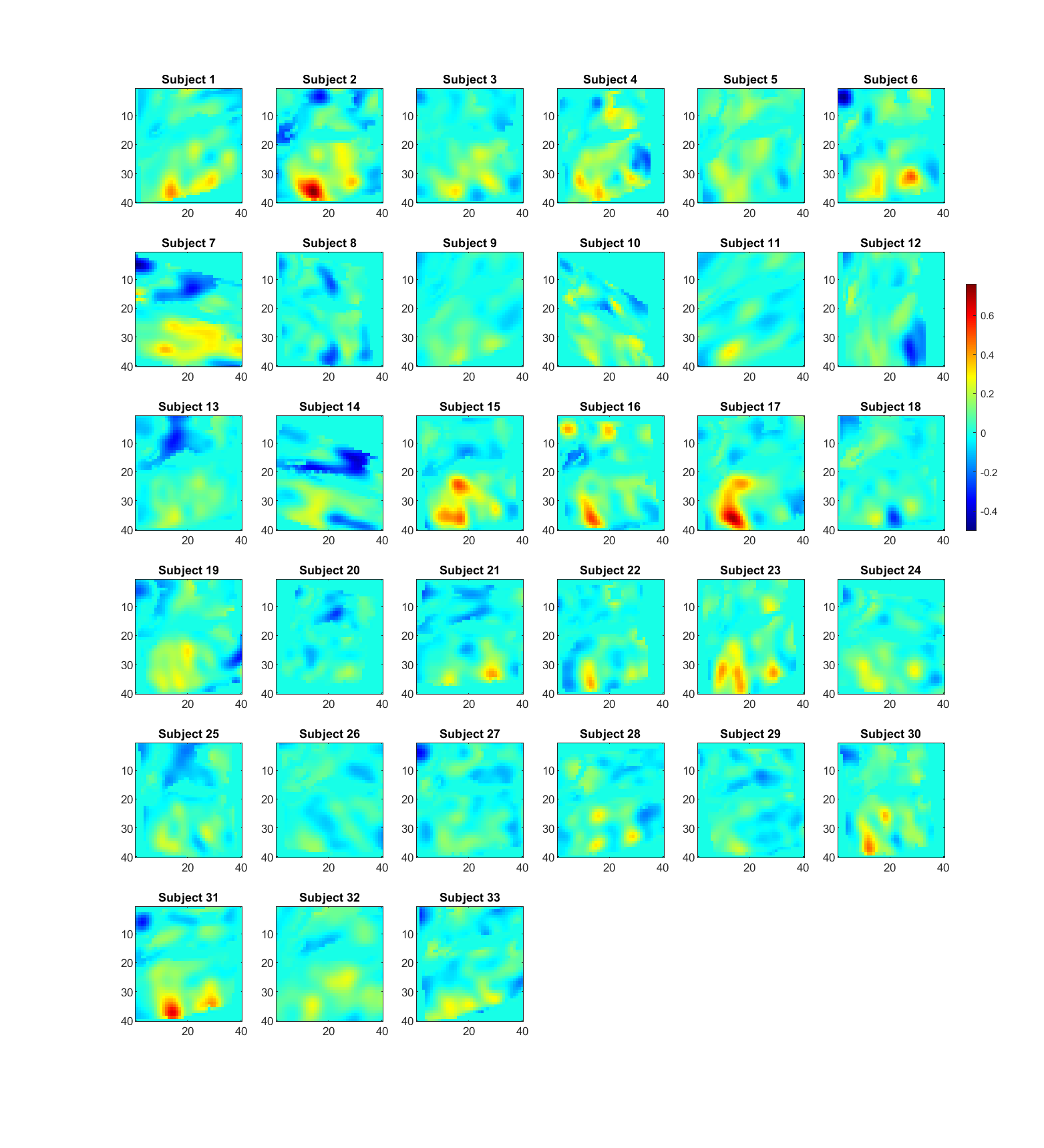}
    \caption{Deformed subject-specific mean activation maps for $33$ subjects.}
    \label{fig: real, inversely warped maps}
\end{figure}



\begin{figure}
    \centering
    \includegraphics[width=0.6\textwidth]{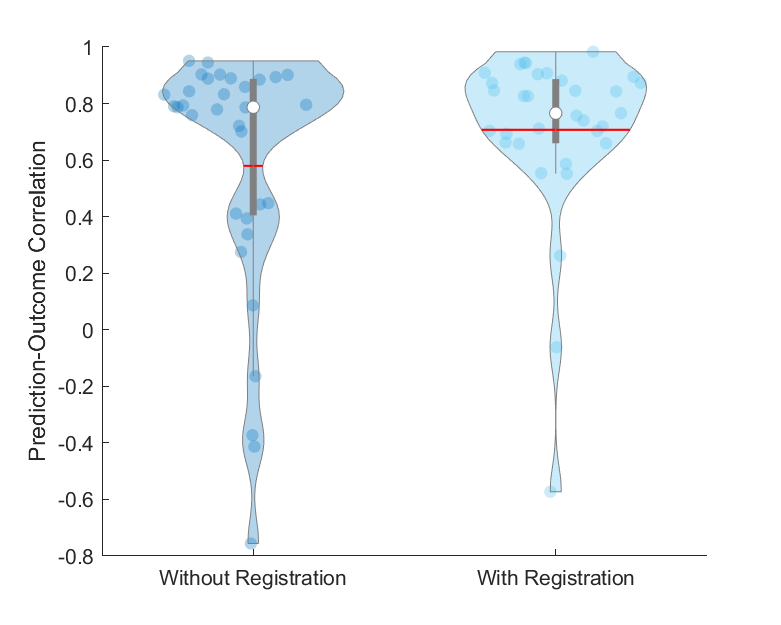}
    \caption{Comparison of the correlations between the predicted pain scores and the true reported pain scores, using the activation maps without registration (left) and with registration (right). The mean correlation (indicated by the red horizontal lines) using the unwarped maps is $0.5796$, compared with $0.7070$ for the warped maps. }
    \label{fig: real, correlation plots, violen plot}
\end{figure}

\begin{figure}
    \centering
    \includegraphics[width=0.6\textwidth]{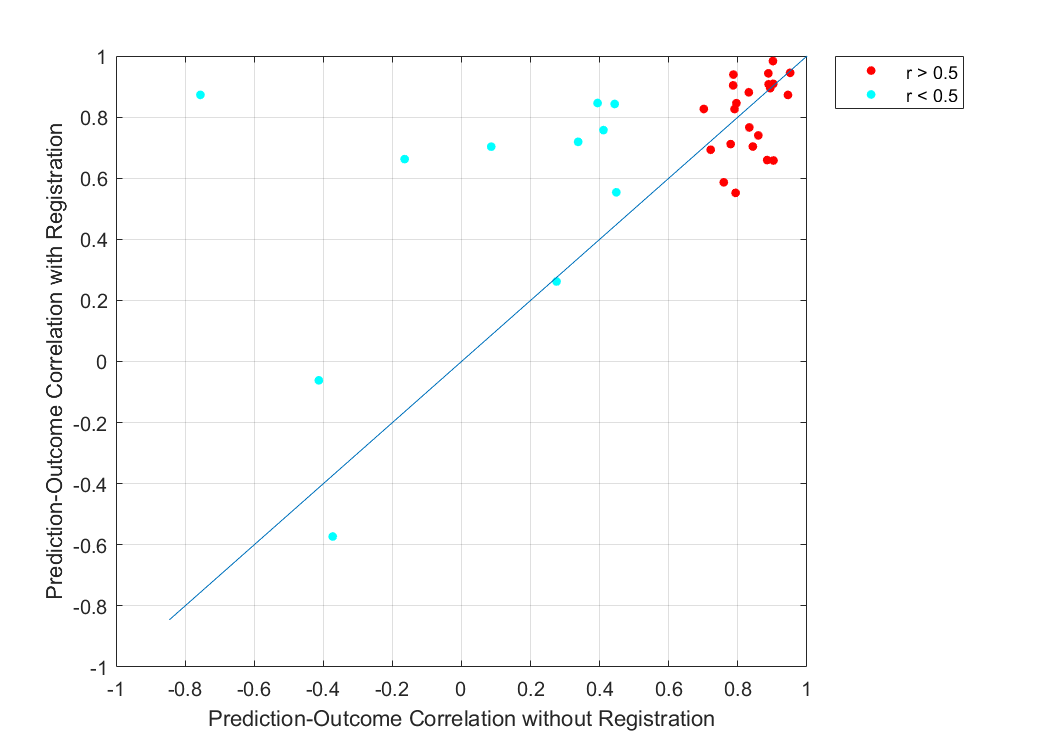}
    \caption{Comparison of the correlations between the predicted pain scores and the true reported pain scores, using the activation maps without and with registration. A $45$ degree line is overlaid for reference (indicting the same values using both approaches). Subjects were grouped into two sets using the prediction-outcome correlations obtained using the model without registration using a threshold of $r=0.5$. In the group with $r<0.5$ (colored in cyan), $9$ out of $11$ subjects ($81.82\%$) showed an improvement in correlation using the proposed approach, while among the rest of subjects (colored in red), $11$ out of $22$ subjects ($50\%$) showed improvement.}
    \label{fig: real, correlation plots, scatter plot}
\end{figure}

\pagebreak
\begin{center}
\textbf{\large Supplemental Materials}
\end{center}
\setcounter{equation}{0}
\setcounter{figure}{0}
\setcounter{table}{0}
\setcounter{section}{0}

\makeatletter
\renewcommand{\theequation}{S\arabic{equation}}
\renewcommand{\thefigure}{S\arabic{figure}}
\renewcommand{\thetable}{S\arabic{table}}
\renewcommand{\thesection}{S\arabic{section}}

\renewcommand{\bibnumfmt}[1]{[S#1]}
\renewcommand{\citenumfont}[1]{S#1}

\section{Updates on the Spatial Parameters $\alpha$ and $\rho$}
Assuming the inverse-Gamma prior assigned to $\alpha$ in (\ref{prior:alpha}), we can express its full conditional distribution as follows:

\begin{small}
\begin{equation}
\begin{aligned}
    \alpha^{k}|\cdot \sim \text{Inv-Gamma}(&a_{0\alpha} + \frac{V(N+1)}{2},\\ 
    &b_{0\alpha} + \frac{1}{2} \bigg(\sum_{i=1}^N\sum_{l=1}^V\frac{(X(T_i(s_l))-X(N_{T_i(s_l)}))^2}{\hat F_{T_i(s_l)}} +\sum_{l=1}^V \frac{(X(s_l)-B_{s_l}X(N(s_l)))^2}{\hat F_{s_l}}\bigg). 
\end{aligned}
\end{equation}
\end{small}
where $\hat F_{s_l} = \hat C(s_l, s_l)- \hat B_{s_l}\hat C_{N_{s_l}}\hat B_{s_l}^T$,
$B_{s_l} =\hat C_{s_l, N_{s_l}}\hat C_{N_{s_l}}^{-1}$, and $\hat C(s_l, s_m) = \exp (-\rho\|s_l-s_m\|)$. The terms $\hat C_{s_l, N_{s_l}}$ and $\hat C_{N_{s_l}}$ are the covariance between $X(s_l)$ and $X(N(s_l))$ and covariance matrix  of $X(N(s_l))$ respectively, with $\alpha=1$. 

The spatial decay parameter $\rho$ is sampled using random walk Metropolis steps with target density (\ref{eqn: symmetric posterior}),
\begin{equation}
\begin{aligned}
    &\prod_{i=1}^N p(Y_i| X(T_i)) p(X(T_i)|X, T_i)  p(X)  p(\rho), \\
    &\approx\prod_{i=1}^N p(Y_i| X(T_i)) \prod_{i=1}^N\prod_{l=1}^V p(X(T_i(s_l))| X(N_{T_i(s_l)})))  \prod_{l=1}^V p(X(s_l)|X(N_{s_l})  p(\rho),
\end{aligned}
\end{equation}
where $p(\rho)$ is the uniform distribution defined in (\ref{prior: rho}).

\section{Updates on $\beta_i$ and $\sigma_i^2$}\label{section: updates beta and sigma}
The posterior distributions of $\beta_i$ and $\sigma_i^2$ also have the closed-form representations, 
\begin{align}
    \beta_i |\cdot &\sim N(\mu_n, \lambda_n),\\
    \sigma_i^2 | \cdot &\sim \text{Inv-Gamma}(a_0+V, a_1+(Y_i^TY_i +Y_i(T_i^r)^TY_i(T_i^r)+ \mu_0^2\lambda_0 -\mu_n^2\lambda_n)/2),
\end{align}
where $\mu_n = \lambda_n(\mu_0\lambda_0 + X(T_i)^T Y_i+ X^TY(T_i^r))$, $\lambda_n = (X(T_i)^TX(T_i)+X^TX+\lambda_0)^{-1}$. We draw samples from the posterior distributions.


\section{Initialization of MCMC Sampling}
The existing group-wise registration algorithms, for example, the iterative approach of \cite{joshi2004unbiased}, can be used to initialize the deformation, $T_i$, and the template $X$. In each iteration, we estimate $T_i$ by warping $\beta_i X$ to $Y_i$ and update $X$ by averaging $Y(T_i^{-1})/\beta_i$. The scaling-correction coefficient $\beta_i$ is estimated as the linear regression coefficient of $Y_i$ onto $X(T_i)$ without the intercept. In each iteration, $T_i$'s and $\beta_i$'s are standardized at identity and $1$ respectively, as described in sections \ref{section: update T_i} and \ref{section: updates beta and sigma}
.

\end{document}